\documentclass{emulateapj}

\usepackage{apjfonts}
\usepackage{amsmath}
\usepackage{amssymb}
\usepackage{natbib}
\usepackage{xspace}
\usepackage{graphicx}
\usepackage{rotate} 
\usepackage{subfigure}
\usepackage{float}

\newcommand{\err}[2]{\ensuremath{^{+#1}_{-#2}}\xspace}
\newcommand{\e}[1]{\ensuremath{^{#1}}\xspace}

\newcommand{\NH}{$N_{\text{H}}$\xspace}

\newcommand{\xte}{\textsl{RXTE}\xspace}

\newcommand{\integral}{\textsl{INTEGRAL}\xspace}
\newcommand{\sax}{\textsl{BeppoSAX}\xspace}
\newcommand{\suzaku}{\textsl{Suzaku}\xspace}
\newcommand{\swiftbat}{\textsl{Swift}-BAT\xspace}
\newcommand{\xmm}{\textsl{XMM-Newton}\xspace}
\newcommand{\chandra}{\textsl{Chandra}\xspace}

\newcommand{\pex}{\textsc{pexrav}\xspace}
\newcommand{\eroll}{$E_{\mathrm{roll}}$\xspace}
\newcommand{\etal}{et al.\xspace}
\newcommand{\ka}{K$\alpha$\xspace}
\newcommand{\chidof}{$\chi^{2}$/dof\xspace}
\newcommand{\fluxunits}{erg\,cm$^{-2}$\,s$^{-1}$\xspace}

\shorttitle{\xte Survey of AGN}
\shortauthors{Rivers, Markowitz \& Rothschild}

\begin{document}

\title{Full Spectral Survey of Active Galactic Nuclei in the \textsl{Rossi X-ray Timing Explorer} Archive}

\author{Elizabeth~Rivers\altaffilmark{1}, Alex~Markowitz\altaffilmark{1,2,3}, Richard~Rothschild\altaffilmark{1}}
\affiliation{$^1$University of California, San Diego, Center for Astrophysics and Space Sciences, 9500 Gilman Dr., La Jolla, CA  92093-0424, USA} 
\affiliation{$^2$Dr.\ Karl Remeis Sternwarte, Sternwartstrasse 7, 96049 Bamberg, Germany} 
\affiliation{$^3$Alexander von Humbolt Fellow} 


\begin{abstract}

We have analyzed spectra for all active galactic nuclei in the \textsl{Rossi X-ray Timing Explorer} (\xte) archive.
We present long-term average values of absorption, Fe line equivalent width, Compton reflection 
and photon index, as well as calculating fluxes and luminosities in the 2--10 keV band for 100 AGN 
with sufficient brightness and overall observation time to yield high quality spectral results.
We compare these parameters across the different classifications of Seyferts and blazars.
Our distributions of photon indices for Seyfert 1's and 2's are consistent with the idea that Seyferts share 
a common central engine, however our distributions of Compton reflection hump strengths do not support the classical 
picture of absorption by a torus and reflection off a Compton-thick disk with type depending only on inclination angle.
We conclude that a more complex reflecting geometry such as a combined disk and torus or clumpy torus is likely a more accurate picture of the Compton-thick material.
We find that Compton reflection is present in $\sim\,$85\% of Seyferts and
by comparing Fe line EW's to Compton reflection hump strengths we have found that on average 40\% of the Fe line 
arises in Compton thick material, however this ratio was not consistent from object to object and did not seem to be dependent on optical classification.

\end{abstract}



\section{Introduction}

Active galactic nuclei (AGNs) are some of the most luminous objects in the universe,
frequently outshining their host galaxies.
Historically AGNs have been classified based on their optical and radio characteristics and were originally 
believed to be a variety of completely different objects.  It is now thought that all AGNs share a common 
central engine: an accreting supermassive black hole (SMBH) located at the center of its host galaxy.  
Angle-dependance and obscuration can explain some of the observed differences in these objects 
but are not enough to explain the great variety of AGN properties that have been discovered.

AGNs display a wide variety of observed behaviors, making it difficult to place them neatly into categories.  
The most general distinction between AGN types is jet-dominated versus non jet-dominated.
Blazars, which are radio bright with typically featureless continua, have particularly strong jet components
which happen to be oriented along our line of sight.  Therefore emission from the beamed
jet dominates over disk/coronal emission from the central region of the AGN.
The remaining categories of AGNs, Quasars, Seyferts, radio galaxies and low luminosity AGNs (LLAGNs), may have jets 
which are not beamed toward us or may have no jets at all. 
Quasars are very luminous, distant AGN while LLAGNs are much fainter and detected only nearby;
however according to unification these objects may differ only in their masses and accretion rates.  
Optical observations are used to identify Seyfert 1's and broad line radio galaxies (BLRG's), which display broadened optical
emission lines, versus Seyfert 2's and narrow line radio galaxies (NLRG's), which display only narrow emission lines.
A special subset of Seyfert 1's are the so-called Narrow Line Seyfert 1's (NLSy1's) which display narrower broad emission
lines than typical Seyfert 1's (H$\beta_{\rm FWHM}$\,<\,2000 km\,s$^{-1}$) and are thought to be relatively small SMBH's 
accreting very near the Eddington limit with very steep (i.e.\ soft) X-ray spectra (see, e.g., Pounds \etal 1995; Grupe \etal 1999).

One thing that nearly all AGNs have in common is a strong X-ray component (Elvis \etal 1978).
It is believed that the X-ray power law continuum arises very near the central black hole, either from the accretion 
disk itself or a hot corona surrounding the black hole, or in some cases possibly from the base of a launching jet.  
In addition to the continuum, Seyfert X-ray spectra can show a variety of components:
emission lines (most commonly from Fe), absorption by gas which may be ionized (``warm'') or neutral (``cold''), 
an excess below about 2 keV known as the ``soft excess,'' and Compton reflection peaking around 20--30 keV in the hard X-ray spectrum.  
Seyfert 2's and NLRG's tend to have more absorption in the line of sight to the nucleus than Seyfert 1's (Antonucci \etal 1993), 
giving rise to the idea that we are seeing them ``edge on'' and that their broad line regions (BLR) exist but are hidden from view by 
the Compton-thick dusty torus which is seen in the infrared.  This classical picture of Seyfert 1/2 unification is still up for debate however, 
with evidence to suggest that while they may share a central engine, differences in the geometry of the circumnuclear material including 
the accretion disk, corona, broad line region, and Compton-thick torus lead to the differences in observed characteristics that we see (see, e.g., Ramos Almeida \etal 2011).
Placing tighter constraints on the geometry of this material may help unravel this mystery of Seyfert 1's and 2's.

The \textsl{Rossi X-ray Timing Explorer} (\xte) performed numerous observations of AGNs over its 16-year lifespan
in the energy range from 3 keV to 200 keV.  This energy range is ideal for quantifying the underlying 
continuum parameters as well as measuring the Compton
reflection hump (CRH), Fe \ka emission, and absorption by cold gas with column densities above $\sim10^{22}$ cm$^{-2}$.
We have performed detailed analysis of all AGNs in the \xte archive for which high-quality spectra could be obtained, totaling 100 objects 
in all. For many of these objects, continuous long-term monitoring data provides long-term baselines for quantities 
such as the continuum photon index ($\Gamma$), the Compton reflection strength ($R$) and the column density of 
absorbing material in the line of sight (\NH).
Relating these components gives us a clearer picture of the geometry of the circumnuclear material in 
Seyfert AGNs and allows us to test unification schemes.

This paper is a follow-up to Rivers \etal (2011; hereafter RMR2011) which analyzed 23 AGN spectra with a particular focus on
high quality data from 20--100 keV, finding average values for the CRH strength, Fe line equivalent width ($EW$), and photon index.
This survey also found limited evidence for high energy rollovers below 100 keV (caused by the thermal shape 
of the corona not being able to produce many very high energy photons) 
with only two objects showing a significant rollover and ruling out the presence of a rollover in all but one of the other sources.
We have quadrupled the sample size and include now all AGNs with high enough data quality
to accurately measure $\Gamma$, as well as the CRH and Fe \ka line $EW$ in most cases.   
The energy range for these spectra was from 3.5 keV up to at least 20 keV and as high as 200 keV for some objects.  
This sample will allow us to analyze the spectral components discussed above with an emphasis on bulk properties
of different types of AGN, providing a test of the unified model.
This paper is structured in the following way: Section 2 contains information on the \xte archive and the data 
reduction process, Section 3 details analysis methods, and Section 4 contains a discussion of our findings.


\begin{figure}
 \epsscale{1.1}
\plotone{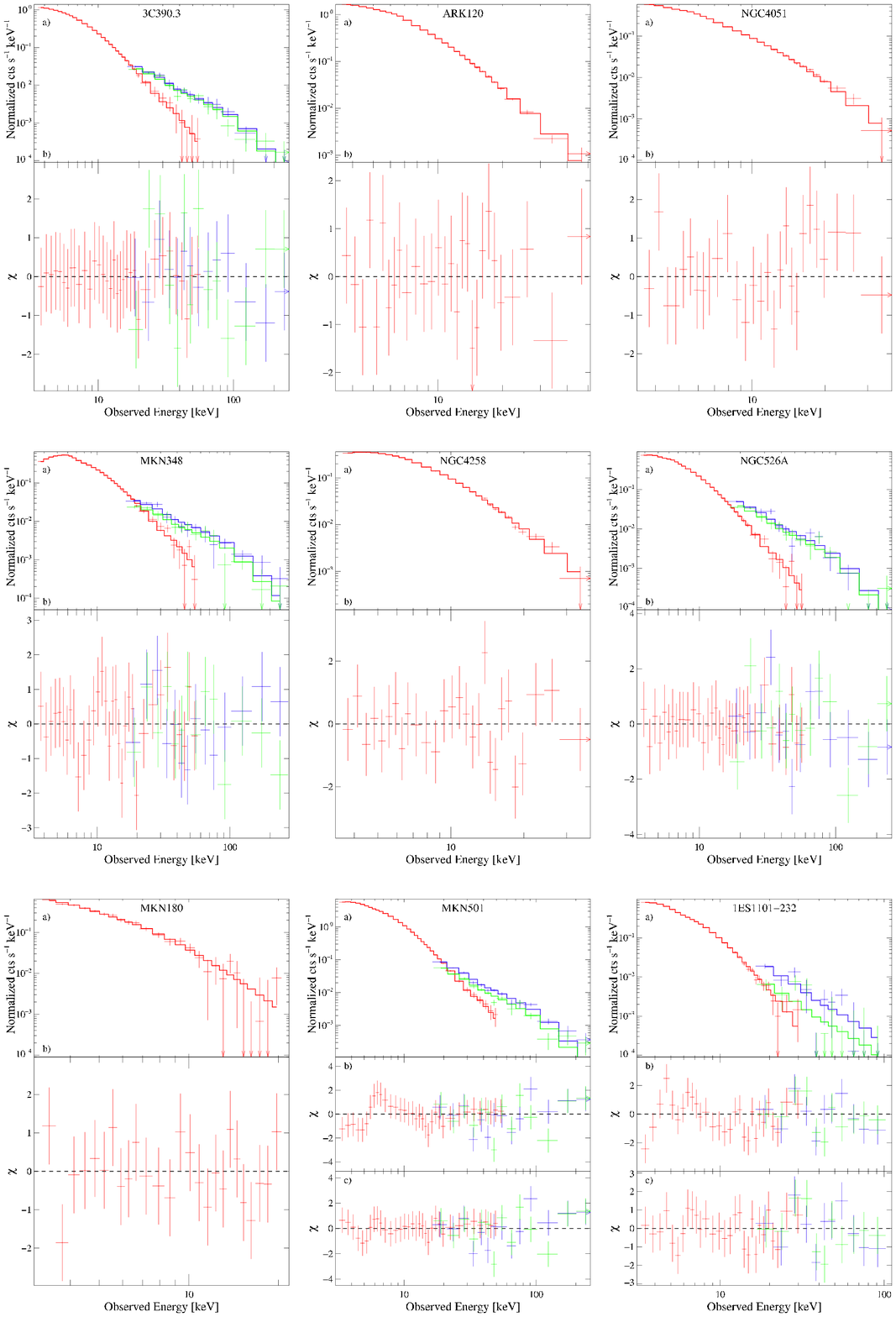}
\caption{Data and data--model fit residuals for a small number of sources in our sample, chosen to display a range of poor to moderate data quality 
	(note that no sources shown here have the highest data quality since those spectra were published in RMR2011).  
	Panel (a) shows the data and best fit model, panel (b) shows the fit residuals to the base model (power law model for blazars), 
	and when present, panel (c) shows the fit residuals to the best fit model (broken power law model for blazars).  Parameters can be found in Tables \ref{tabpar1}--\ref{tabbp}.
	PCA data is in red while HEXTE A and B are blue and green, respectively.  
	The top row are Seyfert 1's, the middle row are Compton-thin Seyfert 2's, and the bottom row are blazars.}   \label{spectra}
\end{figure}

\begin{figure}
\epsscale{1.1}
\plotone{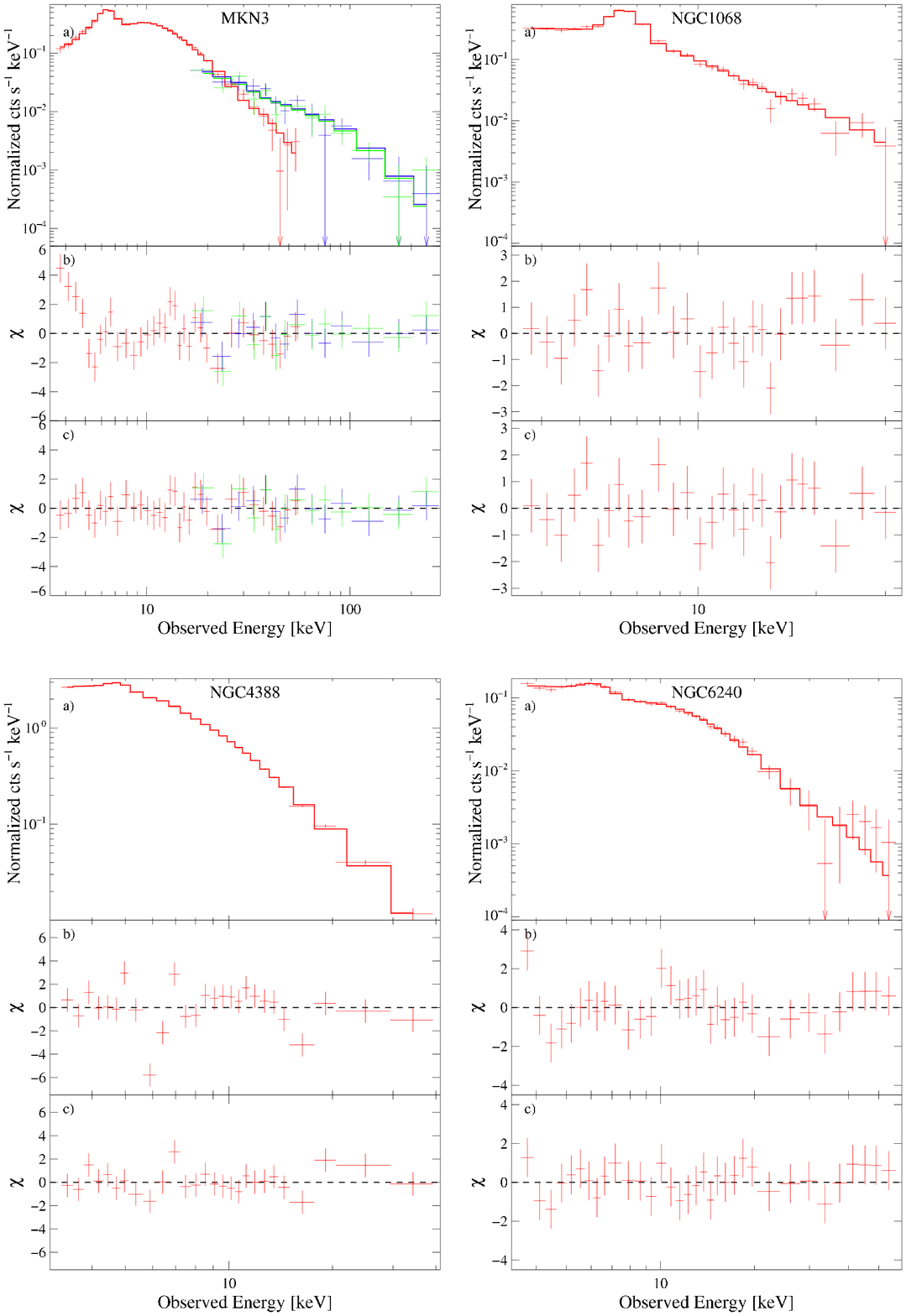}
\caption{Data and data--model fit residuals for four of the complex sources (the other three can be found in RMR2011).  
Panel (a) shows the data and best fit model, (b) shows the fit residuals to the base model from Table \ref{tabpar2}, and (c) shows the fit residuals to the complex model found in Table \ref{tabpc}.
Note that for NGC\,1068 we adopt the base model rather than the complex absorption model based on previous analyses of this source.
}
\label{complex}
\end{figure}

\section{Archival Observations and Data Reduction}

\subsection{The \xte Archive and Selection Criteria}

The \xte satellite made X-ray observations from 1996 January to 2012 January
with its two pointed observation instruments, the Proportional Counter Array 
(PCA; Jahoda \etal 2006) and the High-Energy X-Ray Timing Experiment (HEXTE; Rothschild \etal 1998).
In 16 years it observed 153 AGNs: 54 Seyfert 1's, 47 Seyfert 2's, and 52 Blazars, many of them multiple times.
Note that we have included subtypes Seyfert 1.2 and 1.5 in with the Seyfert 1's and Seyfert 1.8 and 1.9's in with the Seyfert 2's.
The sampling of these objects has been highly inhomogeneous since different viewing schemes 
were proposed for each object at various times and for various scientific goals.  We have included
in our analysis all data for each object, regardless of sampling, in order to construct our
overall averaged spectra.  

We wanted to construct energy spectra of as many of these AGNs as possible, however
several of these object were observed only once or twice for a handful of kiloseconds.  It was
therefore important for us to find the necessary conditions required to construct useful
spectra.  We decided to include a source in our sample if fitting the PCA data with a simple
absorbed power law gave errors on the photon index of $\lesssim\,10\%$.  We included HEXTE data
if the source was detected by HEXTE at the 3$\sigma$ level at 50 keV, otherwise only the PCA data were used.
For the PCA we found that $\sim$\,40,000 total net counts was sufficient to give error bars of $\lesssim\,10\%$ on $\Gamma$.
For HEXTE we found that $\sim$\,5,500 counts were necessary to detect the source at the 3$\sigma$ level at 50 keV,
however the steepness of the spectrum was also a factor in general.
For HEXTE, the whole energy range (20--250 keV) was used whenever the data were included,  though in many cases 
it did not help to constrain the model fitting past $\sim$50--100 keV.  There were two exceptions which had significant 
background features around 100 keV that affected fitting and so were only included to 80--90 keV.

Using these selection criteria we constructed time-averaged spectra for 100 AGN.
We identified these objects by their optical classifications per the NASA/IPAC extragalactic database (``NED''), 
dividing our sample into 34 blazars and 66 Seyferts, including 30 Seyfert 1's, 10 Narrow Line Seyfert 1's, and 26 Seyfert 2's, 7 of which were Compton-thick (defined as \NH $\geq 1 \times 10^{24}$).
The names, redshifts, exposure times, and maximum energy included for each object in our sample are listed in Table \ref{tabsample}.
Those AGN in the archive which did not yield usable spectra are summarized in Appendix A.


\begin{deluxetable*}{ll@{\hspace{0mm}}c@{\hspace{1mm}}c@{\hspace{2mm}}rrr|ll@{\hspace{2mm}}c@{\hspace{2mm}}rrr}
   \tablecaption{Source List \label{tabsample}}
   \tablecolumns{13}
\startdata
\hline
\hline\\[-1mm]
Source Name  &  Type  & W.A. &   $z$   & PCA &  HEXTE   &  E$_{\rm max}$    &  Source Name  &  Type  &   $z$   & PCA &  HEXTE & E$_{\rm max}$  \\[1mm]
&&&& (ks) & A, B (ks) &  (keV)  &&&& (ks) & A, B (ks) &  (keV) \\[1mm]
\hline\\[-1mm]
3C~111  &  BLRG/Sy1  &&  0.0485  &  1092  &  127,~ 278  		& 40/250 	&  Mkn~348  &  Sy2  &  0.0150  &  484  &  68,~  59 		& 60/250  \\
3C~120  &  BLRG/Sy1  &&  0.0330  &  2102  &  504,~ 629  		& 40/250 	&  NGC~526A  &  Sy2/NELG  &  0.0192  &  113  &  35,~  34 & 60/250  \\
3C~382  &  BLRG/Sy1  &&  0.0579  &  154  &  49,~  49 			& 60/250 	&  NGC~1052  &  RLSy2  &  0.0050  &  399  &    		& 40  \\
3C~390.3  &  BLRG/Sy1  &&  0.0561  &  577  &  158,~ 184  		& 60/250 	&  NGC~1068  &  Sy2/C-thick  &  0.0038  &  54  &    	& 30  \\
4U~0241+61  &  Sy1  &&  0.0440  &  152  &  50,~  49  			& 60/250 	&  NGC~2110  &  Sy2/C-thick  &  0.0076  &  194  &  58~  58 & 60/250  \\
Ark~120  &  Sy1  &&  0.0327  &  277  &     						& 50	&  NGC~2992  &  Sy2  &  0.0077  &  70  &    			& 60  \\
Ark~564  &  NLSy1  &&  0.0247  &  447  &     					& 20	&  NGC~4258  &  Sy2/LINER  &  0.0015  &  1463  &    	&  40 \\
Fairall~9  &  Sy1  &&  0.0470  &  647  &     					& 50	&  NGC~4388  &  Sy2/C-thick  &  0.0084  &  98.4  &    	&  40 \\
IC~4329A  &  Sy1  &(1)&  0.0161  &  573  &  148,~ 177  		& 60/100	&  NGC~4507  &  Sy2  &  0.0118  &  145  &  46,~  46 &  60/250 \\
IRAS~13349+2438  &  Sy1/NLSy1  &(2)&  0.1076  & 45 &		& 50	&  NGC~4945  &  Sy2/C-thick  &  0.0019  &  998  &  208,~ 306 & 60/250  \\
MCG--2-58-22  &  Sy1  &&  0.0469  &  223  &  68,~  68   		& 50/250 	&  NGC~5506  &  Sy2  &  0.0062  &  697  &  202,~ 200 	& 60/250  \\
MCG--6-30-15 & NLSy1 & (1) & 0.0077  & 1965 & 505,~ 555  	& 60/250	&  NGC~6240  &  Sy2/C-thick  &  0.0245  &  113  &    	& 55  \\
MCG+8-11-11  &  Sy1  &&  0.0205  &  2  &     					& 60	&  NGC~6300  &  Sy2  &  0.0037  &  27  &    			& 40 \\
Mkn~79  &  Sy1  &&  0.0222  &  1292  &     					& 30	&  NGC~7172  &  Sy2  &  0.0087  &  87  &  26,~  26 		& 50/250  \\
Mkn~110  &  NLSy1  &&  0.0353  &  1283  &     				& 60	&  NGC~7314  &  Sy2  &  0.0048  &  252  &  73,~  73 	& 60/250  \\
Mkn~279  &  Sy1  &&  0.0305  &  180  &     					& 40	&  NGC~7582  &  Sy2/C-thick  &  0.0053  &  185  &  43,~  44 & 60/250  \\
Mkn~335  &  NLSy1  &&  0.0258  &  161  &     					& 25	& 1ES~0229+200  &  BLLAC  &  0.1400  &  279  &    	& 25  \\
Mkn~509  &  Sy1  &&  0.0344  &  738  &  197,~ 224  			& 60/250 	& 1ES~0414+009  &  BLLAC  &  0.2870  &  31  &    		& 20  \\
Mkn~590  &  Sy1  &&  0.0264  &  32  &     					& 40	&  1ES~0647+250  &  BLLAC  &  0.2030  &  42  &    		& 20  \\
Mkn~766  &  NLSy1  &(1) 	&  0.0129  &  771  &     				& 25	&  1ES~1101--232  &  BLLAC  &  0.1860  &  194  &  29,~  29 &  30/250 \\
MR~2251--178  &  Sy1/QSO  &&  0.0640  &  597  &  57,~ 144   	& 50/250 	&  1ES~1218+304  &  BLLAC  &  0.1836  &  10  &    		& 20  \\
NGC~3227  &  Sy1  &(3)&  0.0039  &  1050  &  303,~ 304  		& 60/250	&  1ES~1727+502  &  BLLAC  &  0.0554  &  20  &    		&  20 \\
NGC~3516  &  Sy1  &(4) &  0.0088  &  1036  &  291,~ 290   		& 60/250	&  1ES~1741+196  &  BLLAC  &  0.0840  &  11  &    		& 20  \\
NGC~3783  &  Sy1  &(5)&  0.0097  &  1563  &  204,~ 393  		& 40/80 &  1ES~1959+650  &  BLLAC  &  0.0470  &  229  &  64,~  64 &  50/250  \\
NGC~3998  &  Sy1  &&  0.0035  &  328  &     					& 15	&  1ES~2344+514  &  BLLAC  &  0.0440  &  112  &    	& 40   \\
NGC~4051  &  NLSy1  &(1) &  0.0023  &  1972  &     			& 40	&  1H~0323+342  &  FSRQ  &  0.0610  &  105  &    		& 25   \\
NGC~4151  &  Sy1  &&  0.0033  &  562  &  179,~ 179  			& 60/250	&  3C~273  &  FSRQ  &  0.1583  &  2378  &  429,~ 616  	& 60/250  \\
NGC~4593  &  Sy1  & (6) &  0.0090  &  1389  &  167,~ 326  		& 60/250 &  3C~279  &  FSRQ  &  0.5362  &  2222  &  451,~ 635 	& 25/250  \\
NGC~5548  &  Sy1  & (7)&  0.0172  &  1012  &  294,~312  			& 50/250	 &  3C~454.3  &  FSRQ  &  0.8590  &  54  &  13,~13 		& 40/250  \\
NGC~7213  &  Sy1/Radio  &&  0.0058  &  692  &  				& 25	&  3C~66A  &  BLLAC  &  0.4440  &  162  &    			& 20  \\
NGC~7469  &  Sy1  &&  0.0163  &  1097  &  243,~ 312  			& 20/250	&  4C~29.45  &  FSRQ  &  0.7245  &  159  &    			& 20  \\
PDS~456  &  Sy1/QSO  &&  0.1840  &  361  &     				& 40	&  4C~71.07  &  FSRQ  &  2.1720  &  269  &  0,~  39  	& 25/250  \\
PG~0052+251  &  Sy1  &&  0.1545  &  170  &     				& 40	&  BL~Lac  &  BLLAC  &  0.0686  &  2311  &    			& 30  \\
PG~0804+761  &  Sy1  &(8)&  0.1000  &  382  &     				& 50	&  CTA~102  &  FSRQ  &  1.0370  &  66  &    			& 50  \\
PG~1202+281  &  Sy1  &&  0.1653  &  27  &     				& 25	&  H~1426+428  &  BLLAC  &  0.1291  &  468  &    		&  40 \\
PG~1211+143  &  NLSy1  &&  0.0809  &  131  &     				& 40	&  Mkn~180  &  BLLAC  &  0.0453  &  15  &    			&  20 \\
Pictor~A  &  Sy1/LINER  &&  0.0351  &  34  &     				& 40	&  Mkn~421  &  BLLAC  &  0.0300  &  2230  &  475,~ 481 &  30/250 \\
PKS~0558--504  &  NLSy1  &&  0.1370  &  932  &     			& 20	&  Mkn~501  &  BLLAC  &  0.0336  &  728  &  153,~ 171   & 60/250  \\
PKS~0921--213  &  FSRQ/Sy1  &&  0.0520  &  92  &     			& 40	&  NRAO~530  &  FSRQ  &  0.9020  &  136  &    		& 20   \\
TONS180  &  NLSy1  &&  0.0620  &  326  &     					& 25	&  PG~1553+113  &  FSRQ  &  0.3600  &  119  &    		& 50   \\
Cen~A  &  NLRG  &&  0.0018  &  913  &  109,~ 197  				& 60/250	&  PKS~0528+134  &  FSRQ  &  2.0600  &  247  &    		& 50  \\
Circinus  &  Sy2/C-thick  &&  0.0014  &  103  &  33,~  32   			& 60/250	&  PKS~0548--322  &  BLLAC  &  0.0690  &  13  &    		& 40  \\
Cygnus~A  &  Sy2/Radio  &&  0.0561  &  72  &     				& 40	&  PKS~0829+046  &  FSRQ  &  0.1737  &  240  &    		& 40  \\
ESO~103-G35  &  Sy2  &&  0.0133  &  163  &  50,~  49 			& 60/250	&  PKS~1510--089  &  FSRQ  &  0.3600  &  2091  &  260,~ 387 & 60/250  \\
IC~5063  &  Sy2  &&  0.0113  &  69  &     						& 35	&  PKS~1622--297  &  FSRQ  &  0.8150  &  123  &    		& 50  \\
IRAS~04575--7537  &  Sy2  &&  0.0181  &  49  &     				& 50	&  PKS~2005--489  &  BLLAC  &  0.0710  &  400  &  96,~ 105 & 50/250  \\
IRAS~18325--5926  &  Sy2  &&  0.0202  &  332  &     			& 40	&  PKS~2126--158  &  FSRQ  &  3.2680  &  34  &    		& 20  \\
MCG--2-40-4  &  Sy2  &&  0.0252  &  3  &     					& 25	&  PKS~2155--304  &  BLLAC  &  0.1160  &  902  &    	& 40  \\
MCG--5-23-16  &  Sy2/NELG  &&  0.0085  &  180  &  55,~  54  		& 60/250	&  RGB~J0710+591  &  BLLAC  &  0.1250  &  16  &    	& 25   \\
Mkn~3  &  Sy2  &&  0.0135  &  54  &  15,~  15  					& 60/250	&  S5~0716+714  &  BLLAC  &  0.3000  &  656  &    		& 15  \\\enddata
\tablecomments{Characteristics of our sample.  Sources are listed alphabetically within three groups: Seyfert 1's, Seyfert 2's, and Blazars.  
``BLRG,'' ``NLRG,'' or ``Radio'' indicates a radio loud non-blazar. ``C-thick'' indicates a known Compton-thick object.  
``NELG'' indicates a Narrow Emission Line Galaxy.  Source types were taken from NED.
References for warm absorber parameters are (1) McKernan \etal (2007), (2) Blustin \etal (2005), (3) Markowitz \etal (2009), Turner \etal (2008),
(5) Netzer \etal (2003), (6) Steenbrugge \etal (2003), (7) Steenbrugge \etal (2005) and (8) Pounds \etal (2003).
HEXTE-A and B exposure times are given only for objects where those data were included in our spectral analysis. 
E$_{\rm max}$ is the approximate maximum energy used in our spectral analysis for the PCA/HEXTE.  
For HEXTE, the whole energy range (20--250 keV) was used whenever the data were included,  though in many cases it did not help to constrain the model fitting past $\sim$50--100 keV.  There were two exceptions which had significant background features around 100 keV that affected fitting and so were only included to 80--90 keV.}
\end{deluxetable*}

\subsection{Data Reduction}

For all PCA and HEXTE data extraction and analysis we used HEASOFT version 6.7 software.
Reduction of the data followed standard extraction and screening procedures as detailed in RMR2011.
We used updated PCA background model files ``pca\_bkgd\_cmvle\_eMv20111129.mdl"  for source fluxes brighter 
than $\sim$5 mCrab, and ``pca\_bkgd\_cmfaintl7\_eMv20111129.mdl" for source fluxes fainter than $\sim$5 mCrab.

We extracted PCA STANDARD-2 data from PCU's 0, 1 and 2 prior to 1998 December 23; PCU's
0 and 2 from 1998 December 23 until 2000 May 12; and PCU 2 only after 2000 May 12; using
only events from the top Xe layer in order to maximize signal-to-noise.
Standard screening was applied with time since SAA passage $>$20 minutes and appropriate background models based on brightness 
were selected for each observation provided by the instrument team.
Systematics up to 0.5\% were included for objects with very long exposure times
in order to try to get reduced $\chi^2$ values between 1 and 2 in the best-fit model.

We also obtained HEXTE cluster A and B data for every object, though in many cases there was not
sufficient detection to merit analysis of the HEXTE spectrum (see above).  We did not combine HEXTE A
and B data, and no HEXTE A data were used from after 2006 March when the cluster lost rocking (and therefore
background gathering) capability.  Background subtraction was performed separately for 16 s and 32 s rocking
modes to eliminate problems with differences in the ratio of the on-source and off-source times (see RMR2011 for details).  
Standard HEXTE response matrices were used in all cases.



\begin{deluxetable*}{lrcccccccr}
   \tablecaption{Seyfert 1's: Base Model Fit Parameters\label{tabpar1}}
   \tablecolumns{10}
\startdata
\hline
\hline\\
Source Name  & Flux$_{2-10}$\tablenotemark{A} &  Log(L$_{2-10}$)  &  $\Gamma$ & \NH ~($10^{22}$cm$^{-2}$) &  EW$_{\rm Fe}$ (eV)  &  I$_{\rm Fe}$\tablenotemark{B} &  $R$ &  $FR$  & $\chi^2$/dof \\[1mm]
\hline\\
3C\,111  & 49.1\,$\pm\,$  0.4		& 42.90  &   1.75\,$\pm\,$0.02  &   &     90\,$\pm\,$   20 &	  5.1\,$\pm\,$1.0&	$\leq\,$ 0.1 &    $\leq\,$ 0.1 &   39/50 \\[1mm]
3C\,120  & 37.9\,$\pm\,$  0.7  		& 43.45  &   1.88\,$\pm\,$0.03  &   &    190\,$\pm\,$   70 &    	  8.0\,$\pm\,$3.0&	0.17\,$\pm\,$0.07 &    0.2\,$\pm\,$0.1 &   42/50 \\[1mm]
3C\,382  & 44.4\,$\pm\,$  0.4  		& 44.00  &   1.86\,$\pm\,$0.04  &   &    105\,$\pm\,$   50 &    	  5.5\,$\pm\,$2.6&	0.13\,$\pm\,$0.10 &    0.1\,$\pm\,$0.1 &   26/54 \\[1mm]
3C\,390.3  & 29.4\,$\pm\,$  0.2  	& 43.80  &   1.76\,$\pm\,$0.04  &   &    100\,$\pm\,$   70 &    	  3.6\,$\pm\,$2.4&	0.2\,$\pm\,$0.1 &    0.2\,$\pm\,$0.1 &   31/54 \\[1mm]
4U\,0241+61  & 34.5\,$\pm\,$  0.2  	& 43.65  &   1.74\,$\pm\,$0.04  &   &    200\,$\pm\,$   40 &      8.0\,$\pm\,$1.4&	0.3\,$\pm\,$0.1 &    0.3\,$\pm\,$0.2 &   67/54 \\[1mm]
Ark\,120  & 34.5\,$\pm\,$  0.3  	& 43.40  &   2.07\,$\pm\,$0.05  &   &    240\,$\pm\,$   40 &    	  8.4\,$\pm\,$1.4&	0.5\,$\pm\,$0.2 &    0.5\,$\pm\,$0.2 &    17/29 \\[1mm]
Ark\,564  & 18.5\,$\pm\,$  0.3  	& 42.88  &   2.69\,$\pm\,$0.04  &   &    220\,$\pm\,$  120 &  	  3.2\,$\pm\,$1.7&	- &  - &    22/20 \\[1mm]
Fairall\,9  & 17.7\,$\pm\,$  0.2  	& 43.42  &   2.00\,$\pm\,$0.07  &   &    180\,$\pm\,$   50 &    	  3.4\,$\pm\,$1.0&	0.5\,$\pm\,$0.3 &    0.4\,$\pm\,$0.3 &    27/23 \\[1mm]
IC\,4329A  &102.6\,$\pm\,$  0.4  		& 43.25  &   1.95\,$\pm\,$0.02  &  * &     100\,$\pm\,$   20 &    	 11.2\,$\pm\,$2.1&	0.4\,$\pm\,$0.1 &    0.3\,$\pm\,$0.1 &   69/48 \\[1mm]
IRAS\,13349+2438  &  4.0\,$\pm\,$ 0.2  	& 43.51  &   2.27\,$\pm\,$0.19  & * & 460\,$\pm\,$  460 &  	  2.0\,$\pm\,$2.0&	- &  - &   14/27 \\[1mm]
MCG--2-58-22  & 25.7\,$\pm\,$  0.2  	& 43.58  &   1.70\,$\pm\,$0.04  &   &    160\,$\pm\,$   30 &     4.9\,$\pm\,$1.0& 	$\leq\,$ 0.2 &    $\leq\,$ 0.2 &   47/54 \\[1mm]
MCG--6-30-15  & 41.6\,$\pm\,$  0.3  	& 42.22  &   2.25\,$\pm\,$0.05  & *  &    190\,$\pm\,$   40 &     8.7\,$\pm\,$2.0& 	1.5\,$\pm\,$0.3 &    1.2\,$\pm\,$0.3 &   45/54 \\[1mm]
MCG+8-11-11  & 53.8\,$\pm\,$  0.9  	& 43.19  &   1.70\,$\pm\,$0.07  &   &    210\,$\pm\,$   60 &     11.9\,$\pm\,$3.2& 	$\leq\,$ 0.3 &    $\leq\,$ 0.3 &   15/27 \\[1mm]
Mkn\,79  & 20.3\,$\pm\,$  0.2  		& 42.83  &   1.90\,$\pm\,$0.07  &   &    200\,$\pm\,$   40 &    	  4.4\,$\pm\,$0.9& 	0.7\,$\pm\,$0.3 &    0.6\,$\pm\,$0.3 &   26/28 \\[1mm]
Mkn\,110  & 30.9\,$\pm\,$  0.2  	& 43.41  &   1.80\,$\pm\,$0.04  &   &     65\,$\pm\,$   20 &    	  2.1\,$\pm\,$0.6& 	0.14\,$\pm\,$0.11 &    0.1\,$\pm\,$0.1 &   34/28 \\[1mm]
Mkn\,279  & 19.2\,$\pm\,$  0.3  	& 43.08  &   1.89\,$\pm\,$0.07  &   &    170\,$\pm\,$   60 &    	  3.7\,$\pm\,$1.2& 	$\leq\,$ 0.5 &    $\leq\,$ 0.4 &   22/22 \\[1mm]
Mkn\,335  & 10.7\,$\pm\,$  0.2  	& 42.68  &   2.11\,$\pm\,$0.06  &   &    190\,$\pm\,$   70 &  	  2.0\,$\pm\,$0.7& 	- &  - &   37/21 \\[1mm]
Mkn\,509  & 39.5\,$\pm\,$  0.2  	& 43.50  &   1.87\,$\pm\,$0.03  &   &     80\,$\pm\,$   20 &    	  3.6\,$\pm\,$1.1& 	0.2\,$\pm\,$0.1 &    0.2\,$\pm\,$0.1 &   46/54 \\[1mm]
Mkn\,590  & 33.4\,$\pm\,$  0.5  	& 43.20  &   1.75\,$\pm\,$0.08  &   &    130\,$\pm\,$   70 &    	  4.9\,$\pm\,$2.4& 	$\leq\,$ 0.5 &    $\leq\,$ 0.5 &   17/27 \\[1mm]
Mkn\,766  & 27.8\,$\pm\,$  0.1  	& 42.51  &   2.33\,$\pm\,$0.09  & *  &   1900\,$\pm\,$  950 &    	  1.6\,$\pm\,$0.8& 	0.9\,$\pm\,$0.4 &    0.7\,$\pm\,$0.3 &   26/20 \\[1mm]
MR\,2251--178  & 39.1\,$\pm\,$  0.3  	& 44.03  &   1.76\,$\pm\,$0.01  &   &   60\,$\pm\,$  40 &    	  2.6\,$\pm\,$1.7& 	$\leq\,$ 0.03 &    $\leq\,$ 0.03 &   42/53 \\[1mm]
NGC\,3227  & 32.2\,$\pm\,$  0.2  	& 41.52  &   1.80\,$\pm\,$0.03  & *  &    160\,$\pm\,$   30 &    	  5.8\,$\pm\,$1.2& 	0.3\,$\pm\,$0.1 &    0.3\,$\pm\,$0.1 &   33/54 \\[1mm]
NGC\,3516  & 36.2\,$\pm\,$  0.3  	& 42.28  &   1.85\,$\pm\,$0.04  &  * &    160\,$\pm\,$   40 &    	  7.1\,$\pm\,$2.0& 	0.8\,$\pm\,$0.2 &    0.6\,$\pm\,$0.1 &   43/53 \\[1mm]
NGC\,3783  & 60.9\,$\pm\,$  1.0  	& 42.59  &   1.89\,$\pm\,$0.04  &  * &    290\,$\pm\,$   80 &    	 20.0\,$\pm\,$5.4& 	0.3\,$\pm\,$0.1 &    0.3\,$\pm\,$0.1 &   52/42 \\[1mm]
NGC\,3998  &  7.7\,$\pm\,$  1.0  	& 40.80  &   2.04\,$\pm\,$0.27  &   &    $\leq\,$ 390 &    	   $\leq\,$15.8  & 	$\leq\,$ 1.1 &    $\leq\,$ 0.9 &    18/14 \\[1mm]
NGC\,4051  & 21.1\,$\pm\,$  1.8  	& 40.88  &   2.30\,$\pm\,$0.08  & *  &    140\,$\pm\,$   40 &    	  2.9\,$\pm\,$0.8& 	2.0\,$\pm\,$0.8 &    1.6\,$\pm\,$0.7 &   22/22 \\[1mm]
NGC\,4151  &164.4\,$\pm\,$  0.9  	& 42.39  &   1.88\,$\pm\,$0.01  & 21.2\,$\pm\,$1.0 & 118\,$\pm\,$88 & 39.6\,$\pm\,$29.&    	1.3\,$\pm\,$0.1 &    1.2\,$\pm\,$0.1 &  100/61\tablenotemark{C} \\[1mm]
NGC\,4593  & 38.4\,$\pm\,$  0.3  	& 42.32  &   1.85\,$\pm\,$0.03  &  * &    200\,$\pm\,$   30 &    	  8.2\,$\pm\,$1.1&  	0.3\,$\pm\,$0.1 &    0.3\,$\pm\,$0.1 &   51/54 \\[1mm]
NGC\,5548  & 41.1\,$\pm\,$  0.2  	& 42.91  &   1.89\,$\pm\,$0.02  &  * &    105\,$\pm\,$   25 &    	  4.7\,$\pm\,$1.1&  	0.3\,$\pm\,$0.1 &    0.3\,$\pm\,$0.1 &   46/54 \\[1mm]
NGC\,7213  & 18.9\,$\pm\,$  0.7  	& 41.65 & 1.91\,$\pm\,$0.10  & 		  	&    220\,$\pm\,$   45 &    	  4.1\,$\pm\,$0.9&  $\leq$ 0.3 &    $\leq$ 0.3 &   30/19 \\[1mm]
NGC\,7469  & 27.1\,$\pm\,$  0.3  	& 42.69  &   1.94\,$\pm\,$0.05  &   &    140\,$\pm\,$   50 &    	  4.1\,$\pm\,$1.4&  	0.6\,$\pm\,$0.2 &    0.5\,$\pm\,$0.2 &   47/44 \\[1mm]
PDS\,456  &  7.0\,$\pm\,$  0.1  	& 44.23  &   3.52\,$\pm\,$0.10  &   &    580\,$\pm\,$  320 &  	  4.8\,$\pm\,$2.7&	- &  - &   24/22 \\[1mm]
PG\,0052+251  &  7.3\,$\pm\,$  0.1  	& 44.07  &   1.89\,$\pm\,$0.17  &   &    $\leq\,$ 590 &    	   $\leq\,$5.8   &	$\leq\,$ 1.4 &    $\leq\,$ 1.3 &    16/22 \\[1mm]
PG\,0804+761  & 11.2\,$\pm\,$  0.2  	& 43.89  &   2.00\,$\pm\,$0.06  & *  &    120\,$\pm\,$   60 &     1.7\,$\pm\,$0.8& 	$\leq\,$ 0.3 &    $\leq\,$ 0.2 &    21/23 \\[1mm]
PG\,1202+281  &  5.9\,$\pm\,$  0.2  	& 44.04  &   2.10\,$\pm\,$0.14  &   &    200\,$\pm\,$  140 &  	  1.5\,$\pm\,$1.1&	- &  - &    8/21 \\[1mm]
PG\,1211+143  &  5.5\,$\pm\,$  0.1  	& 43.39  &   1.99\,$\pm\,$0.08  &   &    190\,$\pm\,$   90 &  	  1.2\,$\pm\,$0.6&	- &  - &   14/28 \\[1mm]
Pictor\,A  & 19.8\,$\pm\,$  0.4  	& 43.22  &   1.73\,$\pm\,$0.05  &   &    110\,$\pm\,$   60 &    	  2.5\,$\pm\,$1.3&	$\leq\,$ 0.2 &    $\leq\,$ 0.2 &   14/26 \\[1mm]
PKS\,0558--504  & 14.8\,$\pm\,$  0.2  	& 44.27  &   2.20\,$\pm\,$0.07  &   &    $\leq\,$ 105 &    	   $\leq\,$1.9   &	$\leq\,$ 0.5 &    $\leq\,$ 0.4 &   27/19 \\[1mm]
PKS\,0921--213  &  7.8\,$\pm\,$  0.2  	& 43.15  &   1.66\,$\pm\,$0.14  &   &    $\leq\,$ 190 &    	   $\leq\,$1.7   &	$\leq\,$ 0.8 &    $\leq\,$ 0.8 &    8/29 \\[1mm]
TONS180  &  7.4\,$\pm\,$  0.1  	 	& 43.29  &   2.43\,$\pm\,$0.23  &   &    360\,$\pm\,$  300 &    	  2.2\,$\pm\,$1.8&	$\leq\,$ 2.0 &    $\leq\,$ 1.5 &    20/20 \\[1mm]
\enddata
\tablecomments{Best fit parameters for Seyfert 1's with the base model. Listed are the 2--10 keV observed flux, 2--10 keV unabsorbed luminosity, photon index, column density above the Galactic column, 
			equivalent width of the Fe \ka line, the reflection strength as determined by \pex, the flux ratio of the reflection to the continuum in the 15--50 keV range, 
			and \chidof.  The ``-'' symbol indicates a parameter was unconstrained.  The ``*'' symbol indicates that a warm absorber was modeled with fixed 
			parameters, see references in Table \ref{tabsample}.}
\tablenotetext{A}{Flux is in units of $10^{-12}$ \fluxunits.}
\tablenotetext{B}{Fe line intensity is in units of $10^{-5}$ photons cm\e{-2} s\e{-1}}
\tablenotetext{C}{Fit to complex model given in Table \ref{tabpc}.}
\end{deluxetable*}


\begin{deluxetable*}{lrcccccccr}
   \tablecaption{Seyfert 2's: Base Model Fit Parameters\label{tabpar2}}
   \tablecolumns{10}
\startdata
\hline
\hline\\
Source Name  & Flux$_{2-10}$\tablenotemark{A} & Log(L$_{2-10}$)  &  $\Gamma$ & \NH ~($10^{22}$cm$^{-2}$) &  EW$_{\rm Fe}$ (eV)  &  I$_{\rm Fe}$\tablenotemark{B} &  $R$ &  $FR$& \chidof \\[1mm]
\hline\\
Cen\,A  & 280.7\,$\pm\,$  5.5		& 42.05 & 1.84\,$\pm\,$0.01  &   16.2\,$\pm\,$0.3 &     95\,$\pm\,$   10 &    		 50.7\,$\pm\,$5.9&  $\leq$ 0.01 &    $\leq$ 0.04 &  134/52 \\[1mm]
Circinus  & 22.9\,$\pm\,$  0.8  	& 40.66 & 1.57\,$\pm\,$0.03  & 	  	&   1520\,$\pm\,$   30 &     			 39.0\,$\pm\,$0.9&  6.5\,$\pm\,$0.6 &    6.8\,$\pm\,$0.7 &  786/53\tablenotemark{C} \\[1mm]
Cygnus\,A  & 78.7\,$\pm\,$  8.9  	& 44.36 & 2.06\,$\pm\,$0.09  &    7.1\,$\pm\,$2.7 &    370\,$\pm\,$   70 &    		 42.1\,$\pm\,$7.7&  $\leq$ 0.1 &    $\leq$ 0.1 &   20/20 \\[1mm]
ESO\,103-G35  & 22.0\,$\pm\,$  0.3  	& 42.74 & 1.83\,$\pm\,$0.10  &   28.3\,$\pm\,$2.1 &    290\,$\pm\,$   70 &    	 15.8\,$\pm\,$4.1&  0.5\,$\pm\,$0.2 &    0.5\,$\pm\,$0.2 &   44/53 \\[1mm]
IC\,5063  & 12.2\,$\pm\,$  0.2  	& 42.37 & 1.65\,$\pm\,$0.36  &   31.7\,$\pm\,$11.3 &    180\,$\pm\,$  120 &    		  5.8\,$\pm\,$3.8&  $\leq$ 1.0 &    $\leq$ 1.0 &   18/19 \\[1mm]
IRAS\,04575--7537  & 21.7\,$\pm\,$1.4   & 42.72 & 2.48\,$\pm\,$0.22  &    3.6\,$\pm\,$2.6 &    $\leq\,$350 &    	  4.2\,$\pm\,$4.7&  1.5\err{2.9}{0.9} &    1.1\err{2.0}{0.6} &    10/25 \\[1mm]
IRAS\,18325--5926  & 21.5\,$\pm\,$0.2   & 42.78 & 2.71\,$\pm\,$0.23  &   &    820\,$\pm\,$  270 &    	   		 14.3\,$\pm\,$4.6&  4.5\,$\pm\,$3.3 &    3.1\,$\pm\,$2.3 &    20/20 \\[1mm]
MCG--2-40-4  & 17.6\,$\pm\,$  0.7   	& 42.86 & 1.69\,$\pm\,$0.15  &		  	 &    340\,$\pm\,$  220 &    	  6.3\,$\pm\,$4.1&  $\leq$ 0.8 &    $\leq$ 0.8 &   16/28 \\[1mm]
MCG--5-23-16  & 89.4\,$\pm\,$  1.3  	& 42.71 & 1.85\,$\pm\,$0.04  &    3.7\,$\pm\,$0.8 &    140\,$\pm\,$   20 &    	 16.0\,$\pm\,$2.8&  0.3\,$\pm\,$0.1 &    0.3\,$\pm\,$0.1 &   40/52 \\[1mm]
Mkn\,3  &  7.0\,$\pm\,$  1.6   		& 42.75 & 1.43\,$\pm\,$0.15  &   90.6\,$\pm\,$7.3 &    230\,$\pm\,$   70 &      		  9.2\,$\pm\,$2.9&  0.4\,$\pm\,$0.3 &    0.5\,$\pm\,$0.4 &  100/53\tablenotemark{C} \\[1mm]
Mkn\,348  & 11.6\,$\pm\,$  0.2  	& 42.46 & 1.51\,$\pm\,$0.09  &   17.3\,$\pm\,$2.1 &    125\,$\pm\,$   45 &    		  3.1\,$\pm\,$1.1&  0.3\,$\pm\,$0.2 &    0.4\,$\pm\,$0.2 &   45/53 \\[1mm]
NGC\,526A  & 39.5\,$\pm\,$  1.0  	& 43.07 & 1.80\,$\pm\,$0.08  &    4.6\,$\pm\,$1.6 &     90\,$\pm\,$   40 &    	  4.5\,$\pm\,$2.3	&  0.4\,$\pm\,$0.2 &    0.3\,$\pm\,$0.2 &   47/53 \\[1mm]
NGC\,1052  &  5.9\,$\pm\,$  0.1  	& 41.18 & 1.71\,$\pm\,$0.29  &   13.6\,$\pm\,$5.2 &    190\,$\pm\,$   90 &   2.0\,$\pm\,$1.0	&  $\leq$ 1.6  &    $\leq$1.6 &   41/24 \\[1mm]
NGC\,1068  &  7.6\,$\pm\,$  5.6  	& 41.28 & 1.60\,$\pm\,$0.22  &		  	 &   1880\,$\pm\,$  130 &    	16.8\,$\pm\,$1.1		&  $\leq$ 2.0 &    $\leq$ 2.1 &   27/21\tablenotemark{C} \\[1mm]
NGC\,2110  & 37.5\,$\pm\,$  0.6  	& 42.28 & 1.73\,$\pm\,$0.06  &    6.0\,$\pm\,$1.2 &    190\,$\pm\,$   40 &    	9.6\,$\pm\,$2.2		&  $\leq$ 0.2 &    $\leq$ 0.2 &   30/53 \\[1mm]
NGC\,2992  & 22.3\,$\pm\,$  2.0  	& 41.97 & 1.78\,$\pm\,$0.18  &		  	 &    290\,$\pm\,$   90 &    		     7.4\,$\pm\,$2.4  		&  $\leq$ 0.8 &    $\leq$ 0.7 &   12/27 \\[1mm]
NGC\,4258  &  7.7\,$\pm\,$  1.5  	& 40.23 & 1.80\,$\pm\,$0.10  &    8.4\,$\pm\,$2.0 &    $\leq\,$ 250 &    	$\leq\,$3.0	&  $\leq$ 0.4 &    $\leq$ 0.4 &    23/28 \\[1mm]
NGC\,4388  & 45.2\,$\pm\,$  1.8  	& 42.56 & 1.09\,$\pm\,$0.08  & 		  	&    270\,$\pm\,$   30 &    	 15.9\,$\pm\,$1.8		&  $\leq$ 0.1 &    $\leq$ 0.1 &   79/20\tablenotemark{C} \\[1mm]
NGC\,4507  & 14.0\,$\pm\,$  3.0  	& 42.83 & 1.77\,$\pm\,$0.07  &   86.8\,$\pm\,$2.9 &    150\,$\pm\,$   30 &    	11.9\,$\pm\,$2.5     	&  0.3\,$\pm\,$0.1 &    0.3\,$\pm\,$0.1 &   78/52 \\[1mm]
NGC\,4945  &  4.4\,$\pm\,$  0.2  	& 40.97 & 1.16\,$\pm\,$0.02  & 		  	&      570$\,\pm\,$85	&    		   3.5$\,\pm\,$0.5	&  19\,$\pm\,$2 &   17\,$\pm\,$5 & 1488/53\tablenotemark{C} \\[1mm]
NGC\,5506  & 86.4\,$\pm\,$  0.4  	& 42.35 & 1.98\,$\pm\,$0.03  & 		  	&    320\,$\pm\,$   50 &    	 34.5\,$\pm\,$5.2&  0.8\,$\pm\,$0.1 &    0.7\,$\pm\,$0.1 &   84/53 \\[1mm]
NGC\,6240  &  4.1\,$\pm\,$  2.1    	& 42.92 & 1.45\,$\pm\,$0.18  &		   	&    $\leq\,$ 140 &    	       		   $\leq\,$0.8   &  10\err{12}{4} &   10\err{11}{4} &   36/27\tablenotemark{C} \\[1mm]
NGC\,6300  &  6.2\,$\pm\,$  0.1  	& 40.87 & 1.32\,$\pm\,$0.46  &   14.9\,$\pm\,$7.6 &    450\,$\pm\,$  120 &   		  5.1\,$\pm\,$1.4&  $\leq$ 4.3      &     $\leq$ 4.2        &    10/21 \\[1mm]
NGC\,7172  & 15.9\,$\pm\,$  0.6  	& 42.13 & 1.66\,$\pm\,$0.16  &   16.2\,$\pm\,$3.4 &    180\,$\pm\,$  110 &    		  5.7\,$\pm\,$3.5&  $\leq$ 0.4 &    $\leq$ 0.4 &   61/53 \\[1mm]
NGC\,7314  & 34.6\,$\pm\,$  2.1  	& 41.72 & 1.99\,$\pm\,$0.10  &		  	 &    200\,$\pm\,$   80 &      		  7.5\,$\pm\,$3.0&  0.6\,$\pm\,$0.2 &    0.6\,$\pm\,$0.2 &   39/53 \\[1mm]
NGC\,7582  & 10.5\,$\pm\,$  5.4  	& 41.51 & 1.70\,$\pm\,$0.10  &   13.3\,$\pm\,$2.6 &    340\,$\pm\,$   70 &    		  6.7\,$\pm\,$1.4&  2.7\,$\pm\,$0.9 &    2.7\,$\pm\,$0.7 &   38/53 \tablenotemark{C} \\[1mm]
\enddata
\tablecomments{Best fit parameters for Seyfert 2's with the base model.  Listed are the 2--10 keV observed flux, 2--10 keV unabsorbed luminosity, photon index, column density above the Galactic column, 
			equivalent width of the Fe \ka line, the reflection strength as determined by \pex, the flux ratio of the reflection to the continuum in the 15--50 keV range, 
			and \chidof.  The ``-'' symbol indicates a parameter was unconstrained.}
\tablenotetext{A}{Flux is in units of $10^{-12}$ \fluxunits.}
\tablenotetext{B}{Fe line intensity is in units of $10^{-5}$ photons cm\e{-2} s\e{-1}}
\tablenotetext{C}{Fit to complex model given in Table \ref{tabpc}.  Note that we have adopted the fit given in this table for NGC\,1068 (see Appendix B for details).}
\end{deluxetable*}

\begin{deluxetable*}{lccc@{\hspace{2mm}}c@{\hspace{2mm}}c@{\hspace{2mm}}ccccc@{\hspace{2mm}}cr}
   \tablecaption{Complex Models for Seyferts\label{tabpc}}
   \tablecolumns{11}
\startdata
\hline
\hline\\
&  &     &  && Soft \NH  & Hard \NH &   \\[1mm]
Source Name  &  Flux$_{2-10}$\tablenotemark{A} &  $\Gamma_{\rm hard}$     &  $\Gamma_{\rm soft}$ &   $A_{\rm soft}/A_{\rm hard}$\tablenotemark{B}  &  ($10^{22}$cm$^{-2}$) & ($10^{22}$cm$^{-2}$)  & EW$_{\rm Fe}$ (eV)  & I$_{\rm Fe}$\tablenotemark{C} &  $R$ & \eroll & $\chi^2$/dof \\[1mm]
\hline\\
NGC\,4151   &    175$\,\pm\,$24    &  1.90$\,\pm\,$0.02 & * & 0.44$\,\pm\,$0.20  &  $\leq$\,13.6  &     50\err{50}{10}  	&      299\err{ 251}{ 139} 		&  109$\,\pm\,$71	&   1.0$\,\pm\,$0.2  && 67/60\\[1mm]
Circinus    &  23.2$\,\pm\,$2.7 &  1.2$\pm 0.2$        & 2.5$\,\pm\,$0.4       &   1.1$\,\pm\,$0.3 &  & 920\err{120}{150}  & 2400$\,\pm\,$100 & 52$\,\pm\,$1 & 1.1$\,\pm\,$0.3  &  41\err{6}{10}   &  44/50 \\[1mm]
Mkn\,3      &    7.6$\,\pm\,$1.7   &  1.44$\,\pm\,$0.10  & *  &     0.05$\,\pm\,$0.01 &           &    130\err{ 4}{12}  	&       564\err{ 141}{ 141} 			&  17.6$\,\pm\,$4.4	&  $\leq$\, 0.19  &&     44/52\\[1mm]
NGC\,1068   &    7.6$\,\pm\,$  5.6   &  1.53$\,\pm\,$0.14  & *   &    0.3$\,\pm\,$0.3 &        &   $\geq$780 &    $\leq$\, 4302 			&  57\err{160}{33} 	&  $\leq$\, 1  &    & 22/20\\[1mm]
NGC\,4388  &   48$\,\pm\,$17   &  1.40$\,\pm\,$0.13 & *  &   0.5$\,\pm\,$0.1   &    &     92\err{    8}{   21}  &   257\err{  37}{  67} 			&  22.8$\,\pm\,$4.6 	&  $\leq$\, 0.20  & &    25/19\\[1mm]
NGC\,4945   &  5.1$\,\pm\,$1.2    & 0.88\,$\pm\,$0.12  &    2.06\err{0.6}{0.1}  &  0.8$\,\pm\,$0.1    &   &  425\,$\pm\,$25 &  1420\,$\pm\,$120 &  6.0$\,\pm\,$0.5  & $\leq$ 0.1 &  59\,$\pm\,$7  & 39/51 \\[1mm]
NGC\,6240   &    4.3$\,\pm\,$2.2   &  1.65$\,\pm\,$0.45  &  * &  0.17\err{0.13}{0.04}  &         &    200\err{10}{80}  &     139\err{  32}{ 139} 		&  5.9$\,\pm\,$3.6	&  $\leq$\, 2.90  &&    22/28\\[1mm]
NGC\,7582   &   10$\,\pm\,$6   &  1.75$\,\pm\,$0.12  & *  &   0.69$\,\pm\,$0.15  &  16.5$\,\pm\,$3.9  &   230\err{150}{120}  &     324\err{74}{89}	&  9.4$\,\pm\,$2.4 	&     1.3\err{0.6}{0.4} & &  38/51\\[1mm]
\enddata
\tablecomments{Best fit parameters for Seyferts requiring complex modeling.  For all sources the ratio of the soft to hard power law components are given.
			This may indicate partial covering absorption, scattered emission, or contamination from extended emission in the host galaxy.  
			The hard power law is assumed to be entirely due to AGN activity.
			$\Gamma_{\rm soft}$ was tied to  $\Gamma_{\rm hard}$ (indicated by the "*" symbol) unless it was a significant improvement to leave it free (this was the case in only one source, 
			NGC\,4945, which has significant starburst activity in the host galaxy).  NGC\,4945 and Circinus also required high energy rollovers modeled by \textsc{cutoffpl} 
			with a rollover energy, \eroll, defined as the energy at which the continuum is $1/e$ times the initial value.}
\tablenotetext{A}{Flux is in units of $10^{-12}$ \fluxunits.}
\tablenotetext{B}{The ratio of the normalization at 1 keV of the soft power law to that of the hard power law.}
\tablenotetext{C}{Fe line intensity is in units of $10^{-5}$ photons cm\e{-2} s\e{-1}.}
\end{deluxetable*}

\section{Methods and Analysis}

All spectral fitting for this analysis was done using \textsc{xspec} version 12.5.1k 
with cross-sections from Verner \etal (1996) and solar abundances from Wilms \etal (2000). 
We adopt a standard cosmology of H$_{0}=73.0$, $\Omega_{\Lambda}=0.73$, and $\Omega_{\rm matter}=0.27$.
Uncertainties were calculated at the 90\% confidence level ($\Delta \chi^2$ = 2.71 for one interesting parameter)
unless otherwise stated.

In our model fitting we included free renormalization constants for HEXTE-A and B with respect to the PCA.
This accounts for cross-instrument calibration as well as differences in observing epochs for the different instruments.  
Additionally we included the \textsc{recorn} component in all our models which renormalizes the background level to account 
for slight imperfections in background estimation.  The adjustment was usually less than $\sim$2\% percent.
Bandpasses used for each object were determined on an ad hoc basis, excluding data above a certain energy if
the background dominated the signal. For most objects the range used was 3.5--50 keV, though some of the
very faint or very steep objects (NLSy1's for example) only had usable data up to $\sim$20--30 keV and some of the objects had
high quality HEXTE data up to 100 keV and higher (RMR2011).

For all Seyferts our most basic model included an absorbed power law continuum plus Fe \ka emission modeled by a simple Gaussian 
and the Compton reflection hump modeled with a disk geometry by \pex (Magdziarz \& Zdziarski 1995) which assumes a lamppost-like 
source above a near-infinite plane.  Normalization and photon index of the incident power law in the \pex model were tied to those of 
the continuum power law, abundances were set to solar, and the inclination angle (cos$i$) was frozen at 0.866 (30$\degr$), leaving 
only the reflection fraction $R$ as a free parameter.  See the discussion section for details on this model, its implications, and drawbacks.
Galactic absorption was included for all sources (Kalberla \etal 2005) using the \textsc{phabs} model in \textsc{xspec}.
Warm absorbers were included where well-determined values were found in the literature and had the potential to affect the spectrum 
curvature above 3 keV (i.e., greater than 3\% deviation in the spectrum), using an XSTAR table  component, 
keeping the parameters frozen at the column density and ionization specified in the literature (see Table \ref{tabsample}).
Given the energy range and resolution of the PCA, our data were not sensitive to discrete lines from ionized absorption, 
however rollover from strong, mildly ionized absorbers could be detected below $\sim$\,5 keV.
Additional cold absorption in the line of sight was included for many Seyfert 2's, however for most Seyfert 1's and some Seyfert 2's 
cold absorption in addition to the Galactic column did not cause a significant change in $\chi^2$ and was not included in the base model.
Our best fit values for $\Gamma$, \NH (the column density of cold material in addition to the Galactic column), 
$R$, and the Fe line equivalent width ($EW$) are given in Tables \ref{tabpar1} and \ref{tabpar2} for Seyfert 1's and 2's respectively.
A subsample of spectra are shown in Figure \ref{spectra} to give an idea of the range of data quality in the sample.

The distribution of reduced $\chi^2$ values is fairly smooth with an average value of $\sim$\,1.  
From the average number of degrees of freedom in our sources we would expect a spread of roughly 0.25; instead 
we find a spread of almost twice that (the standard deviation is 0.46).  At the high end this is likely due to systematic errors in very long observations.  
At the low end, there is a known issue with the data reduction software's estimation of background errors.  The software models the background counts
spectrum based on multiple, long blank-sky observations, and then assumes Poissonian errors for the background counts spectrum appropriate for
the exposure time of the observation of the target.  However, the unmodeled residuals in the background are on the order of 1--2$\%$ (Jahoda \etal 2006).  
For sources with a total exposure less than $\sim10-30$ ks, the Poissonian errors may be e.g., 2--4$\%$ of the counts, an overestimate
of the true background errors, and this can in turn yield final errors  on the net (background-subtracted) spectrum which are overestimates
(Nandra et al.\ 2000).  This explains why many of the sources with  relatively short exposures have best-fit models with values of
$\chi^2_{\rm r}$ near 0.6-0.7 (and in these cases, assumption of 1.5$\%$ background errors would yield net spectrum errors that are
smaller by $\sim5-15 \%$, yielding values of $\chi^2_{\rm r} \,\sim\,10-30\%$ higher).  We note that in these cases, because the errors are 
overestimates, our estimates of the errors on best-fit model  parameters reported in the Tables are conservative.
Additionally, for very faint sources ($F_{2-10} \,\sim\, 4-9 \times 10^{-12}$ erg cm$^{-2}$ s$^{-2}$) the average uncertainty in counts/channel could be as high as $\sim$10--20$\%$, 
even when exposure times were over 30 ks (e.g., IRAS~13349+2438, PG~1211+143, PKS~0921--213, NRAO~530, and PKS~0528+134), yielding similarly low values of $\chi^2_{\rm red}$.

In a few cases more complex models were required, specifically either partial covering absorption or scattered nuclear emission
were necessary in a handful of sources. Best fit parameters for these models are shown in Table \ref{tabpc} and spectra for those 
sources not included in RMR2011 are shown in Figure 2.  The majority of these sources are Compton-thick Seyfert 2's, with the exception of NGC\,4151.
Details on these sources can be found in Appendix B.
Note for may of these Compton-thick sources it can be difficult to accurately constrain the parameters
of the coronal power-law component, since there can be degeneracy between $\Gamma$ and the parameters of the CRH and the absorber (see, e.g., RMR2011).
An additional complication is the presence of excess soft emission (below $\sim\,$10 keV), usually modeled as a power law.
This "contaminating" emission can arise from nuclear emission scattered in a diffuse, extended plasma, unresolved point sources in the
host galaxy, starburst activity in the host galaxy, or any combination of these.

Our basic model for the blazars in our sample was a simple power law.  The best fit values for $\Gamma$
and the power law normalization are listed in Table \ref{tabparb}.
We also tried a broken power law model for all blazars but it was only
a significant improvement in fit for four objects, 1ES\,1101--232, 1ES\,1959+650, Mkn\,421 and Mkn\,501.
Best fit parameters for the broken power law model for these objects are given in Table \ref{tabbp}.

\begin{deluxetable*}{lccccr}
   \tablecaption{Blazars: Power-Law Model Parameters\label{tabparb}}
   \tablecolumns{6}
\startdata
\hline
\hline\\
Source Name  &  Flux$_{2-10}$\tablenotemark{A} & Log(L$_{2-10}$) &  $\Gamma$   & $A$ $(10^{-2})$ &   $\chi^2$/dof  \\[1mm]
\hline\\
1ES\,0229+200  & 15.1\,$\pm\,$0.8	& 44.69  &  1.88\,$\pm\,$0.03  &   0.49\,$\pm\,$0.02  &   37/25 \\[1mm]
1ES\,0414+009  &  8.8\,$\pm\,$1.9  	& 43.82  &  2.68\,$\pm\,$0.12  &   0.90\,$\pm\,$0.19  &   14/22 \\[1mm]
1ES\,0647+250  & 17.8\,$\pm\,$2.4  	& 45.03  &  2.67\,$\pm\,$0.08  &   1.79\,$\pm\,$0.25  &   19/21 \\[1mm]
1ES\,1101--232  & 40.7\,$\pm\,$1.3  	& 44.43  &  2.51\,$\pm\,$0.02  &   3.29\,$\pm\,$0.10  &   74/54\tablenotemark{B} \\[1mm]
1ES\,1218+304  & 12.1\err{2}{6}  	& 43.29  &  2.53\,$\pm\,$0.20  &   1.01\err{0.17}{0.47}  &    8/16 \\[1mm]
1ES\,1727+502  &  9.5\,$\pm\,$1.4  	& 44.30  &  2.00\,$\pm\,$0.07  &   0.37\,$\pm\,$0.05  &   11/23 \\[1mm]
1ES\,1741+196  & 19.1\,$\pm\,$3.3  	& 44.69  &  2.15\,$\pm\,$0.10  &   0.93\,$\pm\,$0.16  &    6/17 \\[1mm]
1ES\,1959+650  &148.3\,$\pm\,$1.9  	& 44.97  &  2.12\,$\pm\,$0.01  &   6.83\,$\pm\,$0.09  &   71/59\tablenotemark{B} \\[1mm]
1ES\,2344+514  & 25.0\,$\pm\,$1.2  	& 43.95  &  2.15\,$\pm\,$0.03  &   1.21\,$\pm\,$0.06  &   30/28 \\[1mm]
1H\,0323+342  & 15.3\,$\pm\,$1.2  	& 44.34  &  1.88\,$\pm\,$0.04  &   0.50\,$\pm\,$0.04  &   28/24 \\[1mm]
3C\,273  & 98.5\,$\pm\,$1.0  		& 43.51  &  1.70\,$\pm\,$0.00  &   2.43\,$\pm\,$0.02  &   72/60 \\[1mm]
3C\,279  &  9.1\,$\pm\,$0.5  		& 43.58  &  1.68\,$\pm\,$0.03  &   0.22\,$\pm\,$0.01  &  88/51 \\[1mm]
3C\,454.3  & 66.7\,$\pm\,$3.1  		& 45.22  &  1.63\,$\pm\,$0.02  &   1.47\,$\pm\,$0.07  &   28/53 \\[1mm]
3C\,66A  &  6.4\,$\pm\,$1.0  		& 45.24  &  2.73\,$\pm\,$0.10  &   0.70\,$\pm\,$0.11  &   14/15 \\[1mm]
4C\,29.45  &  3.1\,$\pm\,$0.5  		& 46.52  &  1.74\,$\pm\,$0.08  &   0.08\,$\pm\,$0.01  &   27/24 \\[1mm]
4C\,71.07  & 16.0\,$\pm\,$7.7  		& 44.93  &  1.53\,$\pm\,$0.02  &   0.30\,$\pm\,$0.14  &   45/37 \\[1mm]
BL~Lac  & 11.2\,$\pm\,$0.5  		& 46.70  &  1.83\,$\pm\,$0.02  &   0.34\,$\pm\,$0.02  &   36/26 \\[1mm]
CTA\,102  &  9.7\,$\pm\,$1.3  		& 43.55  &  1.81\,$\pm\,$0.07  &   0.28\,$\pm\,$0.04  &   30/28 \\[1mm]
H\,1426+428  & 23.6\,$\pm\,$0.7  	& 45.84  &  1.92\,$\pm\,$0.02  &   0.81\,$\pm\,$0.02  &   18/28 \\[1mm]
Mkn\,180  & 12.3\,$\pm\,$2.7  		& 44.42  &  2.70\,$\pm\,$0.13  &   1.28\,$\pm\,$0.28  &    8/14 \\[1mm]
Mkn\,421  &419.4\,$\pm\,$5.6  		& 43.23  &  2.70\,$\pm\,$0.01  &  43.86\,$\pm\,$0.58  &  283/57\tablenotemark{B} \\[1mm]
Mkn\,501  &109.6\,$\pm\,$1.1  		& 44.40  &  2.00\,$\pm\,$0.01  &   4.26\,$\pm\,$0.04  &   76/59\tablenotemark{B} \\[1mm]
NRAO\,530  &  3.5\,$\pm\,$1.0  		& 43.92  &  2.24\,$\pm\,$0.16  &   0.19\,$\pm\,$0.05  &    12/15 \\[1mm]
PG\,1553+113  & 14.5\,$\pm\,$1.2  	& 45.27  &  2.61\,$\pm\,$0.05  &   1.36\,$\pm\,$0.11  &   18/28 \\[1mm]
PKS\,0528+134  &  4.1\,$\pm\,$0.5  	& 45.10  &  1.65\,$\pm\,$0.06  &   0.09\,$\pm\,$0.01  &   21/28 \\[1mm]
PKS\,0548--322  & 32.9\,$\pm\,$3.0  	& 46.06  &  2.15\,$\pm\,$0.05  &   1.59\,$\pm\,$0.15  &   12/28 \\[1mm]
PKS\,0829+046  &  3.3\,$\pm\,$0.8  	& 44.02  &  2.11\,$\pm\,$0.14  &   0.15\,$\pm\,$0.03  &   22/17 \\[1mm]
PKS\,1510--089  &  6.7\,$\pm\,$0.4  	& 44.76  &  1.35\,$\pm\,$0.03  &   0.09\,$\pm\,$0.01  &  111/59 \\[1mm]
PKS\,1622--297  &  8.4\,$\pm\,$0.6  	& 45.57  &  2.07\,$\pm\,$0.04  &   0.36\,$\pm\,$0.03  &   29/28 \\[1mm]
PKS\,2005--489  & 56.0\,$\pm\,$0.9  	& 44.27  &  2.46\,$\pm\,$0.01  &   4.25\,$\pm\,$0.07  &   40/54 \\[1mm]
PKS\,2126--158  &  8.4\,$\pm\,$1.0  	& 46.78  &  1.66\,$\pm\,$0.07  &   0.19\,$\pm\,$0.02  &   12/25 \\[1mm]
PKS\,2155--304  & 33.2\,$\pm\,$0.6  	& 44.47  &  2.68\,$\pm\,$0.01  &   3.40\,$\pm\,$0.07  &   24/19 \\[1mm]
RGB\,J0710+591  & 40.9\,$\pm\,$3.4  	& 44.63  &  2.18\,$\pm\,$0.05  &   2.07\,$\pm\,$0.17  &   13/24 \\[1mm]
S5\,0716+714  &  3.9\,$\pm\,$0.8  	& 44.37  &  2.51\,$\pm\,$0.11  &   0.32\,$\pm\,$0.06  &   17/13 \\[1mm]

\enddata
\tablecomments{Best fit parameters for blazars with the simple power law model.  Listed are the 2--10 keV flux, the photon index, and the normalization 
			of the power law defined as ph\,keV$^{-1}$\,cm$^{-2}$\,s$^{-1}$ at 1 keV.  Note that 1ES 1218+0304 includes additional systematic errors 
			due to possible contamination by Mkn~766 as detailed in the text.}
\tablenotetext{A}{Flux is in units of $10^{-12}$ \fluxunits.}
\tablenotetext{B}{Better fit by broken power law model given in Table \ref{tabbp}.}
\end{deluxetable*}
\begin{deluxetable*}{lccccccr}
   \tablecaption{Blazars: Broken Power-Law Model Parameters\label{tabbp}}
   \tablecolumns{8}
\startdata
\hline
\hline\\
Source Name  &  Flux$_{2-10}$\tablenotemark{A} & Log(L$_{2-10}$) &  $\Gamma_{1}$   & $\Gamma_{2}$  &   $A$~$(10^{-2})$ &  E$_{\rm break}$ (keV) & $\chi^2$/dof  \\[1mm]
\hline\\
1ES\,1101-232  &   39.3\,$\pm\,$0.6 	 & 44.96	& 2.31\,$\pm\,$0.20  &   2.56\,$\pm\,$0.06  &   2.49\,$\pm\,$0.63  &  4.6\,$\pm\,$ 1.5  &   55/52 \\[1mm]
1ES\,1959+650 &  145.4\,$\pm\,$0.7	 & 44.33	&  1.99\,$\pm\,$0.07  &   2.14\,$\pm\,$0.01  &   5.70\,$\pm\,$0.52  &  4.9\,$\pm\,$ 0.6  &   34/57 \\[1mm]
Mkn\,421  		  &  367\,$\pm\,$7  	 & 44.35	&  2.41\,$\pm\,$0.09  &   2.75\,$\pm\,$0.01  &  26.01\,$\pm\,$3.81  &  6.6\,$\pm\,$ 0.4  &   73/55 \\[1mm]
Mkn\,501  		  & 109.0\,$\pm\,$0.3	 & 43.92	&  1.97\,$\pm\,$0.02  &   2.02\,$\pm\,$0.01  &   4.04\,$\pm\,$0.11  &  6.9\,$\pm\,$ 1.2  &   55/57 \\[1mm]
\enddata
\tablecomments{Best fit parameters for blazars with the broken power law model.  The normalization, $A$ is defined as ph\,keV$^{-1}$\,cm$^{-2}$\,s$^{-1}$ at the break energy.
			All show significant improvement in the fit over a simple power law, though 1ES\,1101-232 has a break energy very close to the edge of the bandpass
			and should be treated with caution.}
\tablenotetext{A}{Flux is in units of $10^{-12}$ \fluxunits.}
\end{deluxetable*}

\begin{figure}
  \plotone{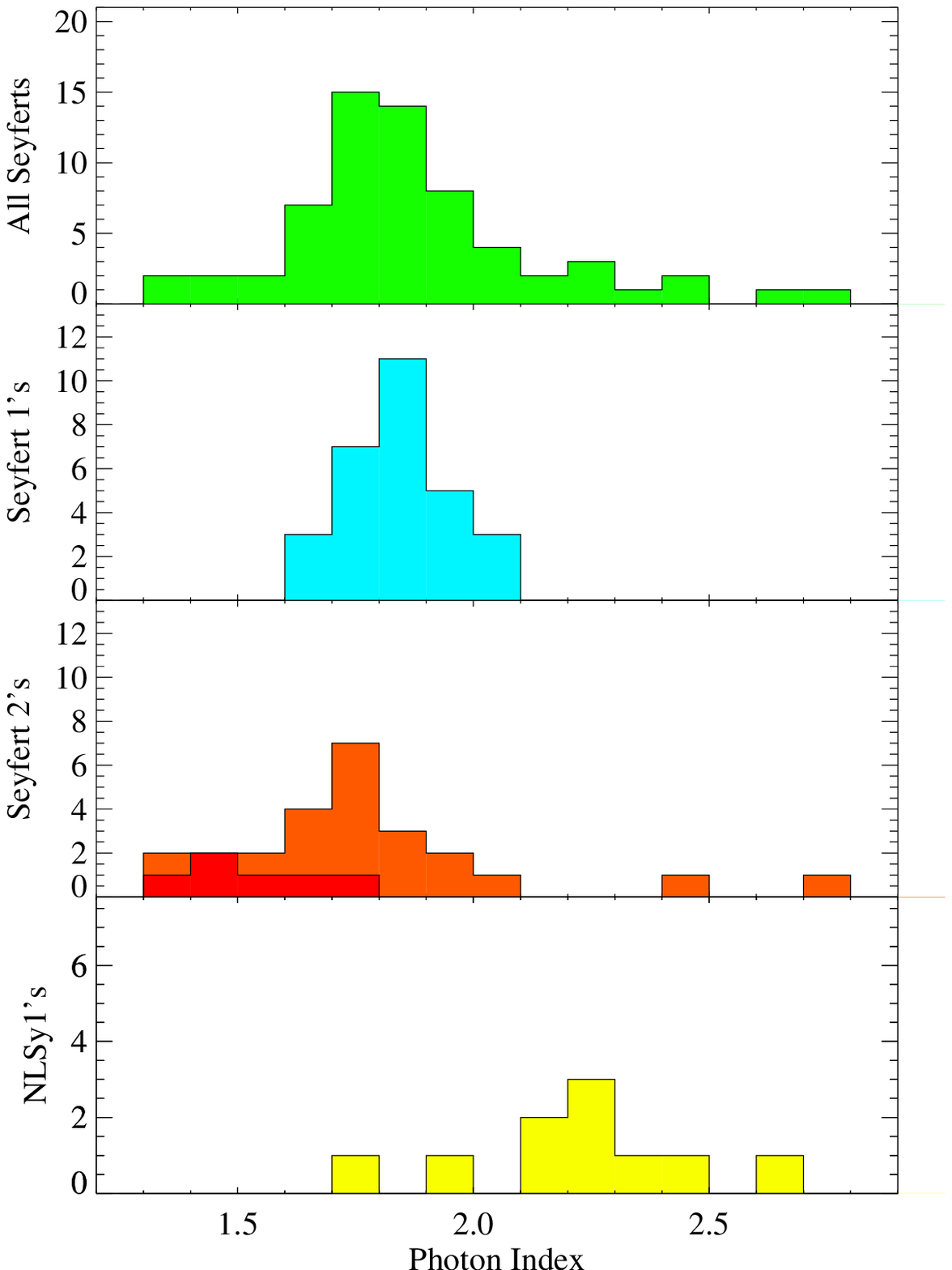}
  \caption{$\Gamma$ distribution by type.  Seyfert 2's are divided into Compton-thin (orange) and Compton-thick (red).  Typical Seyfert 1's and Seyfert 2's are all consistent with an average photon index of 1.8--1.9 as has been found in previous works.  NLSy1's show much steeper X-ray spectra with $\Gamma$\,>\,2 in most sources.  Note that two Seyfert 2's, IRAS\,18325--5926 and IRAS\,04575--7537 have X-ray characteristics similar to NLSy1's, being very steep without absorption $\geq$10\e{22} cm\e{-2} and a poorly constrained CRH.  NGC\,4945 is an extreme outlier and is not shown on this plot.}
  \label{Gdist}
\end{figure}

\begin{figure}
  \plotone{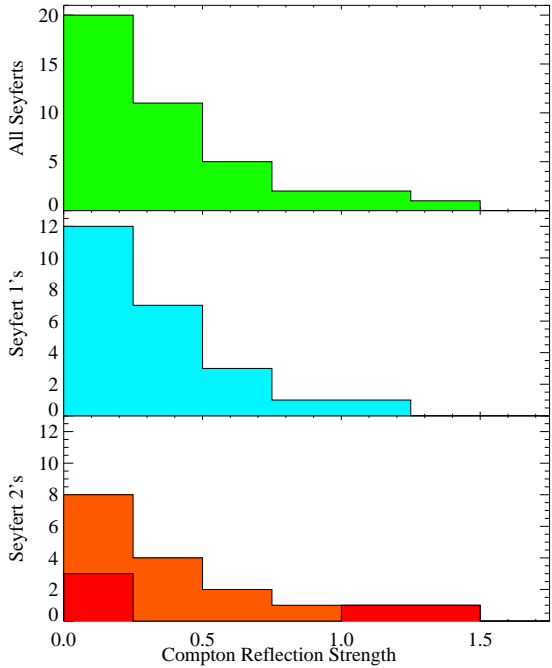}
  \caption{$R$ distribution by type. Seyfert 2's are divided into Compton-thin (orange) and Compton-thick (red).  NLSy1's are not included with the Seyfert 1's.  Sources with only upper limits are all located in the far left bin and sources without well-determined $R$ values (i.e., with $\sigma_{R}/R>1$ and upper limits $\geq$\,0.5) are left off this plot.}
  \label{Rdist}
\end{figure}

\begin{figure}
  \plotone{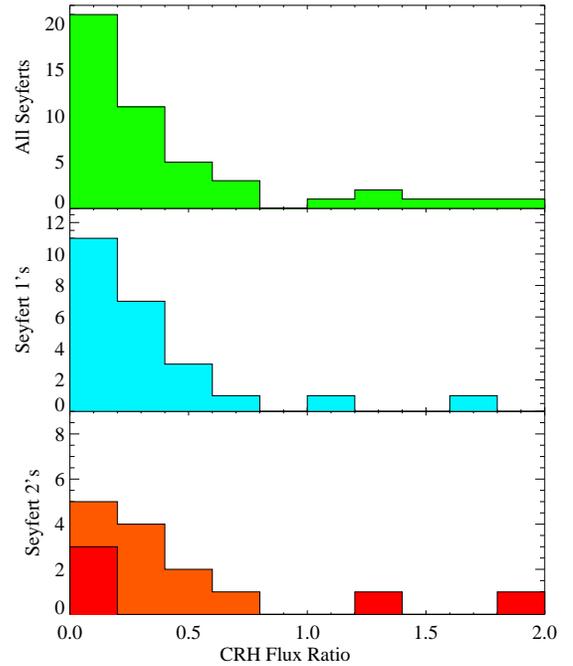}
  \caption{$FR$ distribution by type, where $FR$ is defined as the ratio of the CRH flux to the power law flux in the 15--50 keV range and can be used to make comparisons with other reflection models which are commonly utilized (i.e., $FR$ is model independent).  Seyfert 2's are divided into Compton-thin (orange) and Compton-thick (red).  NLSy1's are not included with the Seyfert 1's.  Sources with only upper limits are located in the far left bin and sources without well-determined $FR$ values (i.e., with $\sigma_{FR}/FR>1$ and upper limits $\geq$\,0.5) are left off this plot.}
  \label{FRdist}
\end{figure}


\begin{deluxetable}{lcc}
   \tablecaption{Average Spectral Parameter Values by Optical Classification\label{tabavg}}
   \tablecolumns{3}
\startdata
\hline
\hline\\[-1mm]
Type                      &  $\Gamma$     &  $R$   \\[1mm]
\hline
\hline\\[-1mm]
All Seyferts              &   1.90      &    0.45    \\[1mm]
\hline\\[-1mm]
Narrow Line Seyfert 1's   &   2.24      &    0.89    \\[1mm]
Seyfert 1's               &   1.86      &    0.27    \\[1mm]
Compton-thin Seyfert 2's  &   1.85      &    0.27    \\[1mm]
Compton-thick Seyfert 2's &   1.40      &    0.48    \\[1mm]
\hline\\[-1mm]
Blazars                   &   2.1         &    \\[1mm]
\hline\\[-1mm]
BLLAC                     &   2.3         &     \\[1mm]
FSRQ                      &   1.8         &     \\[-1mm]
\enddata             
\tablecomments{Average model parameter values for sources in our sample by type.  Objects with poorly constrained parameters have been omitted when calculating these averages.  Note that the high average $R$ value for all Seyferts is due in large part to the contribution from the steep NLSy1's.  For typical Seyfert 1's and 2's the average $R$ is $\sim$0.3.}
\end{deluxetable}

\begin{deluxetable*}{lccccccc}
   \tablecaption{Comparisons to Selected Surveys: Average Spectral Parameter Values\label{tabsurveys}}
   \tablecolumns{8}
\startdata
\hline
\hline\\[-1mm]
                      								&    This Work    & Dadina08  & Patrick12 & Ricci11 & Nandra94  & Gondek96  & Winter09 \\[1mm]
\hline
\hline\\[-1mm]
All Seyferts ........................... $\Gamma$           	  	&   1.90    &    1.8	&      		&		&  1.95  & 	   	&	1.78 \\[1mm]
~~~~~~~~~~~~~~~~~~ ............................ $R$           	&   0.45    &    1.0	&	   	&		&  1.60  &     	&	 \\[1mm]
Seyfert 1's ............................. $\Gamma$           		&   1.86    &    1.89* 	&  1.82	  &	1.96	&  	     & 	   1.90	\\[1mm]
~~~~~~~~~~~~~~~~~~ ............................ $R$           	&    0.27   &    1.23*	&  	  	&	0.2	&  	     & 	   0.76	 \\[1mm]
Narrow Line Seyfert 1's  ........ $\Gamma$     		&   2.24    &		&  	  	&	2.28	&  	  \\[1mm]
~~~~~~~~~~~~~~~~~~ ............................  $R$		&    0.89   &		&  	  	&	4.3	&  	\\[1mm]
Compton-thin Seyfert 2's ...... $\Gamma$     		&   1.85    &    1.80*	&  		  &	1.97	&  	 \\[1mm]         
~~~~~~~~~~~~~~~~~~ ............................ $R$  	  	&    0.27   &    0.87*	&  		  &	2.0	&  	 \\[1mm] 
Compton-thick Seyfert 2's.... $\Gamma$    		&   1.40    &		&  		  &	1.9	&  	 \\[1mm]     
~~~~~~~~~~~~~~~~~~ ............................ $R$           	&    0.48   &		&  		  &	1.4	&  	 \\[1mm]
\enddata             
\tablecomments{Comparing our average spectral parameters to several other surveys of Seyferts in the hard X-ray band.  We find that we are consistent in general with other surveys though a number of specific cases of discrepancies highlight that the high variance among Seyferts means that the makeup of any given sample is important.  In particular Dadina (2008) did not separate out NLSy1's or Compton-thick versus Compton-thin Seyfert 2's.    The energy band and analysis methods can also have a strong influence on measured values of $R$ as demonstrated by our fitted values to stacked spectra versus equally weighted averages (see Section 4.1).  The "*" symbol indicates averages that may not separate NLSy1's from the typical Seyfert 1's or Compton-thick Seyfert 2's from the Compton-thin Seyfert 2's.  The other surveys are Dadina (2008), Patrick \etal (2012), Ricci \etal (2011), Nandra \& Pounds (1994), Gondek \etal (1996), and Winter \etal (2009).}
\end{deluxetable*}


\section{Discussion}

Our excavation of the \xte archive has produced a unique sample of 100 AGNs with spectral data from 
3.5 keV to $\gtrsim$\,20 keV.  The breadth of this energy range has allowed us to explore key 
spectral components that have not been well-studied to date.  Most significantly, quantifying the Compton 
reflection hump requires spectral sensitivity over a broad energy range which many other modern X-ray observatories
lack (\chandra, \xmm, {\it Swift}).  \xte's ability to observe the very hard X-ray properties of AGNs simultaneously
with their mid-range (2--10 keV) X-ray properties eliminates problems associated with non-simultaneous observing
which can be particularly severe in highly variable objects.  
Additionally, \xte does not suffer from cross-calibration uncertainties between instruments such as between the 
\suzaku XIS and HXD or between the \sax MECS and PDS instruments.
Several AGN studies at high X-ray energies ($\gtrsim$\,10keV) have been performed with \sax, \textsl{CGRO}-OSSE, \swiftbat, \integral, and \suzaku 
(Dadina 2007; Zdziarksi \etal 2000; Tueller \etal 2010; Ricci \etal 2011; and Patrick \etal 2012, respectively), 
particularly focusing on Seyferts.  We begin our discussion by presenting the results of our analysis and then 
comparing them to those from other surveys.

\begin{figure}
  \plotone{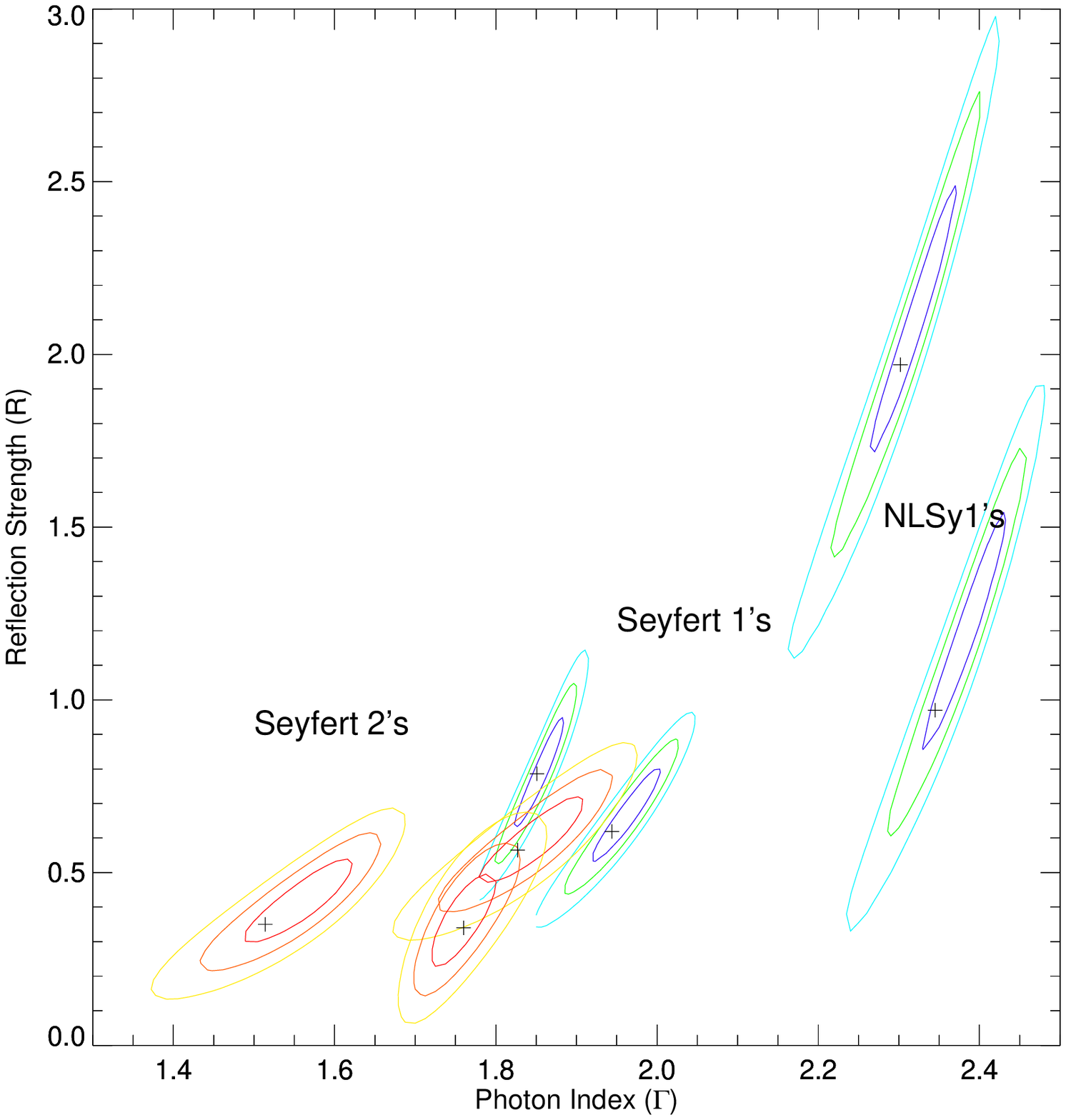}
  \caption{Contours for selected low flux (1--4$\times 10^{-11}$\,\fluxunits) Seyferts showing degeneracy between $R$ and $\Gamma$. 
  	The contour lines going from dark to light (inner to outer) correspond to 3$\sigma$, 2$\sigma$, and 1$\sigma$; blue is used for Seyfert 1's and red is used for Seyfert 2's.
  	 From left to right the sources are Mkn\,348, NGC\,4507, ESO 103-G35, NGC\,3516, NGC\,7469, NGC\,4051, and Mkn\,766. 
           $R$ is particularly difficult to constrain in NLSy1's compared to the other Seyfert classifications due to the steepness of the spectra.}
  \label{figcontours}
\end{figure}

\begin{figure}
  \plotone{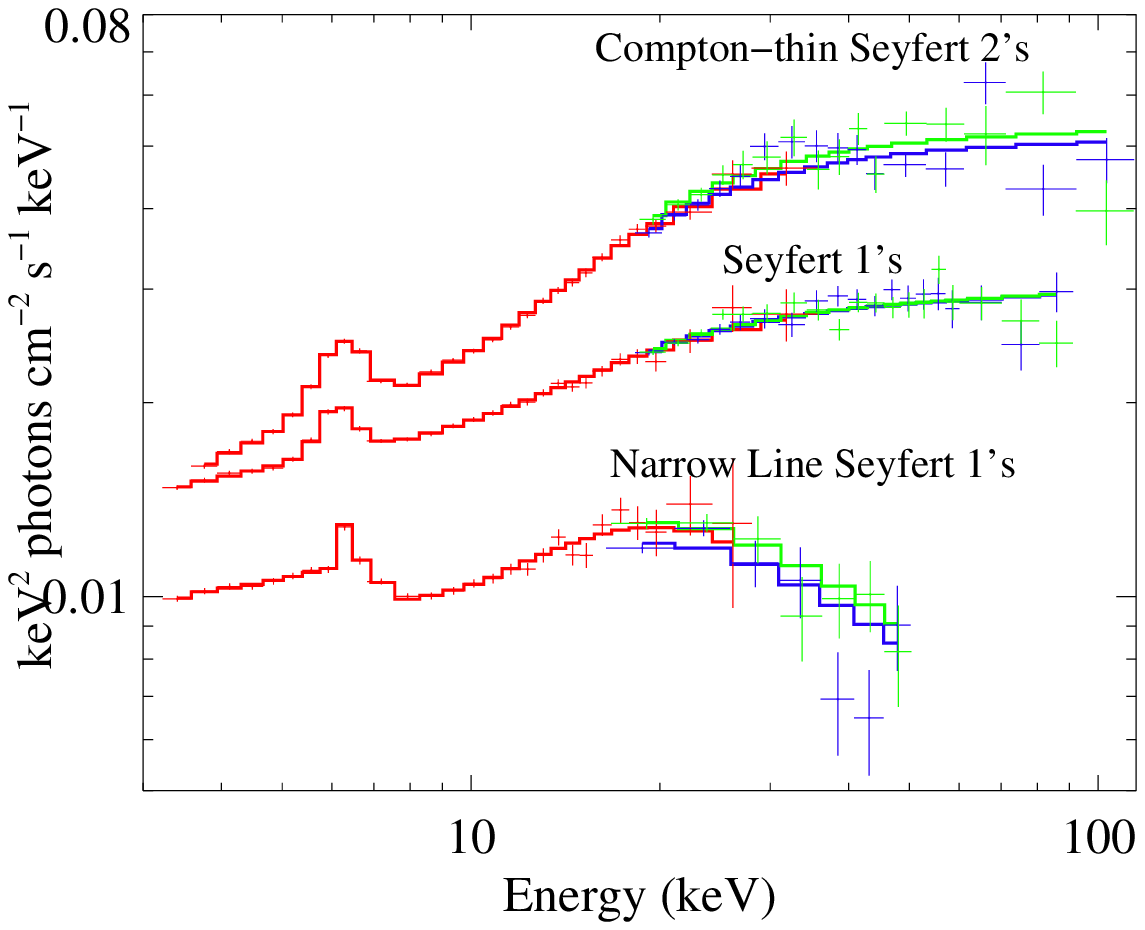}
  \caption{Stacked X-ray $\nu$F$_{\nu}$ spectra for Seyfert 1's (excluding NLSy1's), Compton thin Seyfert 2's (excluding Cen\,A), and Narrow Line Seyfert 1's.  
  	Red, blue and green data points and model lines denote PCA, HEXTE-A, and HEXTE-B.  
	We found best fit photon index values of 1.85$\,\pm\,$0.02, 1.77$\,\pm\,$0.03 and 2.18$\,\pm\,$0.08 for Seyfert 1's, 2's and NLSy1's respectively. 
	Values of $R$ from the stacked spectra were found to be 0.5$\,\pm\,$0.1, 1.1$\,\pm\,$0.1, and 1.5$\,\pm\,$0.7.  
	For the NLSy1's, the peak around 20 keV is the CRH which shows up clearly above the downward sloping power law.}
  \label{SED}
\end{figure}

\subsection{Results for the Seyfert Sample}

Making use of the large sample provided by the \xte archive we can examine spectral properties of different types of AGNs.  
We have divided our sample into optically classified Seyfert 1's, Seyfert 2's, 
NLSy1's, Compton-thick Seyfert 2's and blazars (which will be discussed in a following section).  
Unweighted average parameter values for the various Seyfert sub-types are given in Table \ref{tabavg}.
The distributions of $\Gamma$ and $R$ by object type are given in Figures \ref{Gdist} and \ref{Rdist}.

The average photon index for our entire sample of Seyferts was 1.94.  For Seyfert 1's and Compton-thin Seyfert 2's 
the average photon indices were 1.86 and 1.79 respectively with standard deviations of $\sim$0.12.  
The similarity in $\Gamma$ between these two classes of objects supports the Seyfert 1/2 unification schemes since we 
would expect the intrinsic photon indices to be unrelated to the viewing angle.
They are also consistent with the values of $\sim$1.8--1.9 generally accepted to be the average range of power law photon indices in Seyferts 
(e.g.\  Nandra \& Pounds 1994, Gondek \etal 1996, Dadina 2008).
The Compton-thick Seyfert 2's had an average photon index of 1.77 with a standard deviation of 0.26, although for these sources it is very 
difficult to measure the intrinsic photon index accurately since the extreme curvature of the spectrum gives little leverage for measuring $\Gamma$.

NLSy1's had an average photon index of 2.24 (with a standard deviation of 0.24), significantly higher than other Seyferts and consistent with the idea that 
these objects are in a different regime of accretion (Pounds \etal 1995).  If Seyfert 1's and 2's share a common central engine, we would expect
to see Seyfert 2's with similar properties to NLSy1's which could not be identified optically (since the BLR is obscured in Seyfert 2's).   
IRAS\,18325--5926 and IRAS\,04575--7537 are Seyfert 2's which show very soft power laws with photon indices of 
2.71$\pm$0.23 and 2.48$\pm$0.22, respectively.  These sources resemble NLSy1's in their X-ray spectra, and may be part of a
class of objects that have been a missing piece in the Seyfert 1/2 unification puzzle.
NGC~5506 is has been shown to be a hidden NLSy1 (Nagar \etal 2002) and has a photon index of 1.98$\pm$0.03.

We detected a strong CRH ($R \gtrsim $0.2) significant at the 5$\sigma$ level in 28 of the 66 Seyferts in our sample.
Only 5 showed no contribution from the CRH at all ($R<0.1$).
Thus, of the 33 sources which had enough counts enough to measure $R$ with high significance, $\sim$85\% showed at least some contribution from the CRH.
The remaining 33 Seyferts did not have well measured CRH's due to a lack of counts above 10 keV and/or a weak reflection hump.
Averages were calculated excluding sources with poorly constrained $R$ values, i.e., those with only upper limits that were greater than 0.5.
The average reflection strength for all Seyferts was 0.45 with a standard deviation of 0.76 (note that the distribution is non-Gaussian and highly skewed; see Figure \ref{Rdist}).
Note also that in Compton-thick sources and NLSy1's it is difficult to constrain the level of the power law continuum against which $R$ is measured.  
These sources are likely to have overestimated $R$ values for this reason.
Seyfert 1's had an average $R$ value of 0.27 and Compton-thin Seyfert 2's had an average of 0.27, and standard deviations of 0.28 and 0.27 respectively, 
consistent with Seyfert 1's and 2's having on average the same amount of reflected flux. 
Contour plots of $\Gamma$ versus $R$ for selected Seyferts are shown in Figure \ref{figcontours}.

We have also created stacked spectra for Seyfert 1's, Compton-thin Seyfert 2's and Narrow Line Seyfert 1's, including
all objects weighted by exposure (excluding Cen\,A which dominates the Seyfert 2 stacked spectrum otherwise).
Combining these into an overall X-ray SED with the correct relative abundances of different source types
could give a good idea of the contribution of AGN to the cosmic X-ray background (CXB; see, e.g., Gilli \etal 2007).
The individual spectra used to create the stacked spectra were not put into local reference frames, however
blurring of the Fe line and edge due to our including sources spanning a range of redshifts was less than the energy resolution of the PCA.
These stacked spectra are shown in Figure \ref{SED} as $\nu$F$_{\nu}$ plots, giving the X-ray portion of the spectral energy distribution (SED)
and clearly showing the difference in spectral shape between NLSy1's and other Seyferts.

Results of fitting the base model to these spectra yielded average photon indices of 1.85$\,\pm\,$0.02, 1.77$\,\pm\,$0.03 and 2.18$\,\pm\,$0.08 
for Seyfert 1's, 2's and NLSy1's respectively. Values of $R$ from the stacked spectra were found to be 0.5$\,\pm\,$0.1, 1.1$\,\pm\,$0.1, and 1.5$\,\pm\,$0.7.  
Absorption in the line of sight was not significant to include in any of the sub-sets.
Even though many Seyfert 2's show significant absorption $>$10\e{22} cm\e{-2}, the soft end of the Compton-thin Seyfert 2 stacked 
spectrum seems to be dominated by NGC~5506, which has a relatively high flux, long exposure time, and a very low column density.
Note that these fitted parameters are significantly higher than our average unweighted values in Table \ref{tabavg}.
This demonstrates that weighting is an important factor in CXB synthesis models, even within a given type of object, 
due to the high variation among these sources.  
In any given patch of the sky, the portion of the CXB due to unresolved AGN could vary in spectral shape due to individual sources.  
We will base the remainder of our discussion on our sample averages from Table \ref{tabavg} which give 
equal weight to all objects and are therefore not dominated by the most-observed/brightest sources.

\begin{figure}
  \plotone{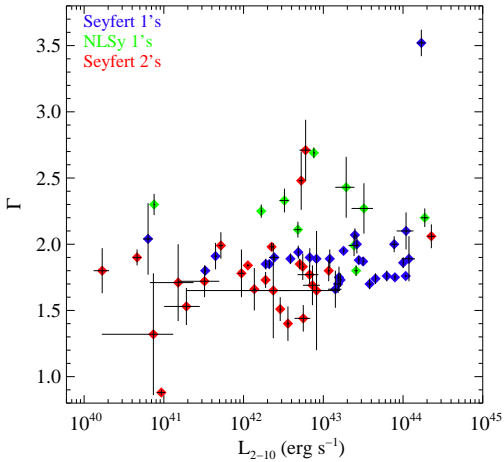}
  \caption{The photon index showed no significant correlation with the 2--10 keV unabsorbed X-ray luminosity. }
  \label{LG}
\end{figure}

\begin{figure}
  \plotone{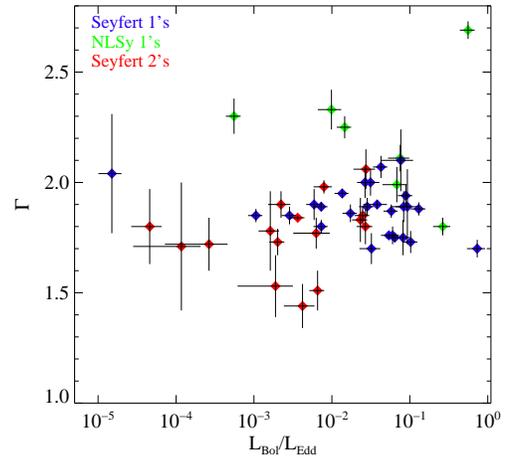}
  \caption{Photon index versus Eddington ratio showed no significant correlation. Black hole masses were obtained for 49 sources primarily from Vestergaard \& Peterson (2006), and when not present there, from Winter \etal (2009) and Merloni \etal (2003).  Converting our X-ray luminosity to bolometric luminosity we used a luminosity-dependent scaling factor from Marconi \etal (2004).}
  \label{LGedd}
\end{figure}

Figures \ref{LG} and \ref{LGedd} show the relationship between luminosity and photon index for our Seyfert sample.
We converted the X-ray luminosity to bolometric luminosity ($L_{\rm Bol}$) using a luminosity-dependent scaling factor from Marconi \etal (2004).
Eddington luminosity ($L_{\rm Edd}$) values were computed from black hole masses for 49 sources, primarily taken from Vestergaard \& Peterson (2006), 
and when not present there, from Winter \etal (2009) and Merloni \etal (2003).  
We do not find a significant correlation between $\Gamma$ and $L_{2-10}$ or $\Gamma$ and  $L_{\rm Bol}/L_{\rm Edd}$.
Sobolewska \& Papadakis (2009) analyzed \xte monitoring data of 10 AGN and
found that for a given AGN, $\Gamma$ is correlated with both $F_{2-10}$ and the mass accretion rate, \textsl{\.m}$_{X,E}$ ($L_{2-10}/L_{\rm Edd}$),
and that across their sample there was a positive correlation between $\Gamma$ and \textsl{\.m}$_{X,E}$.
We do not see this trend across our entire sample of AGN.  This could be due to a number of factors, for instance their objects were all quite bright and 
one of them, Ark~564, lies at one extreme corner of the $\Gamma$ versus Eddington ratio plot in Figure \ref{LGedd}.  
We do not find a correlation between $\Gamma$ and $L_{\rm Bol}/L_{\rm Edd}$ for the other nine objects in their sample 
(note that our black hole masses were not identical to theirs and this happens to lessen the correlation considerably as well).


\begin{figure}
  \epsscale{0.9}
  \plotone{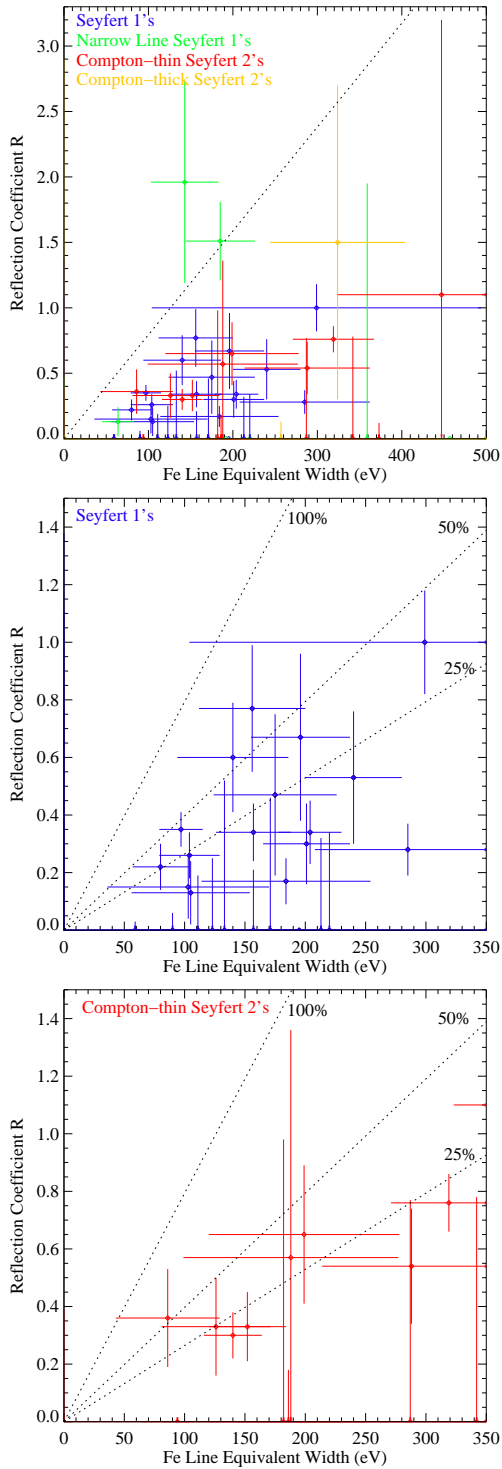}
  \caption{Comparing the amount of reflection, $R$, to the EW of the Fe \ka line for Seyferts.  The left-most dashed line is the predicted amount of Fe emission associated with reflection off Compton-thick material in a disk.  Points falling the right of this line must have additional contribution from Compton thin material.  The dashed lines indicate the relative amount of Fe line flux contributed by Compton-thick material.  Seyfert 1's and Compton-thin Seyfert 2's have been isolated in the bottom panels for clarity.  Note the different axis ranges for these panels and that no sources with $EW > 500$ eV are included here.}
  \label{RvsEW}
\end{figure}

\subsection{Comparisons to Previous Surveys}

A large number of X-ray spectral surveys of AGN have been performed in the past, with anywhere from a small 
handful to over one hundred objects, and with a variety of energy ranges.
Average values of $\Gamma$ and $R$ for several surveys  which included data above 10 keV
are given in Table \ref{tabsurveys}.

Dadina (2008) performed a survey on 105 Seyferts with \sax data in the 2--100 keV range, finding average values of 
$\Gamma \sim\,1.8$, $R \sim\, 1.0$, and \eroll$\sim\,290$ keV.
Their average photon index is lower than ours, likely due to their inclusion of high energy 
rollovers which would tend to cause slightly their slightly flatter values of $\Gamma$.
We did not do an extensive search for high energy rollovers in this sample since previous work has already been done with the 
highest quality HEXTE data (RMR2011) which ruled out the presence of a rollover below 200 keV in most cases.
Their exceptionally high average $R$ value for Seyfert 1's of 1.23 is less easy to explain.
Additionally, Dadina (2008) found correlations between $\Gamma$ and $R$, and between $L_{2-10}$ and the $EW$ of the neutral Fe \ka line, 
the ``X-ray Baldwin Effect," neither of which could be confirmed in our data.  
Again, the inclusion of several Compton-thick sources could have an effect given the degeneracy in measuring $\Gamma$ and $R$ in these sources.

Patrick \etal (2012) analyzed a sample of 46 Seyfert 1's observed with \suzaku and \swiftbat and found that 
39 out of their 46 Seyfert 1's showed a significant CRH, exactly in line with our 85\% of sources.
However since they used a different Compton reflection model we cannot directly compare our $R$ values with their results, 
necessitating an alternative measurement of the strength of the CRH, as detailed in the following section.
They found an average $\Gamma$ for their sample of 1.82$\pm$0.03, in agreement with our value for Seyfert 1's of 1.86$\pm$0.27.
Their inclusion of Seyfert 1.8--1.9 in their sample should not have an effect on this value since we find that there is no difference in 
the average value of $\Gamma$ between Seyfert 1--1.5's and Compton thin Seyfert 1.8-2's.

Other surveys have been performed with more limited energy ranges and/or fewer objects.
A survey of Seyferts at high X-ray energies was done by Ricci \etal (2011) using stacked 
\integral-ISGRI data in the 17--250 keV range for different classes of Seyferts to obtain bulk spectral properties. 
Discrepancies between their results and ours may be due to their lack of coverage below 17 keV, making it difficult to quantify
the underlying power law in sources with strong Compton reflection, modeling of the CRH (assumptions about inclination angle can change measured $R$ values),
and stacking itself, which as we have discussed above can lead to domination by just a few sources. 
They also fitted their spectra with models that included high energy rollovers in the 100--300 keV range.
When they modeled high energy rollovers their average $\Gamma$ and $R$ values for Seyferts 1's were 1.8 and 0.1, for Compton-thin Seyfert 2's 1.6 and 0.4,
for Compton-thick Seyfert 2's 1.9 and 1.4, and for NLSy1's 2.3 and 4.2, demonstrating the degeneracy between $\Gamma$, $R$ and \eroll.

Nandra \& Pounds (1994) analyzed \textsl{Ginga} data of 27 Seyferts in the 1.5--37 keV range and found average $\Gamma$ and $R$ values of 1.95 and 1.60. 
Gondek \etal (1996) produced stacked Seyfert 1 spectra using combined \textsl{CGRO}-OSSE, \textsl{Ginga} and \textsl{EXOSAT} data (taken at different times)
with average values for $\Gamma$ and $R$ of 1.90\,$\pm\,$0.05 and 0.76\,$\pm\,$0.15. 
Both these samples are consistent with our average values for the entire sample of 1.9 and 0.54 respectively.
Winter \etal (2009) found an average $\Gamma$ of 1.78 with a standard deviation of 0.24 for a sample of 102 \swiftbat 
selected AGN, however they did not model any Compton reflection.  

Our \xte results are consistent with previous analyses and our sample has a number of advantages over
previous surveys in the medium--hard X-ray bandpass. 
Since the PCA and HEXTE have always operated simultaneously, we do not have the ambiguity from 
source variability that comes from combining non-simultaneous 
soft and hard X-ray data sets from different missions as is commonly necessary to obtain broadband coverage.
The broad bandpass is necessary to accurately constrain Compton reflection and gain insight into the 
geometry and characteristics of the circumnuclear material.
Additionally, many of these sources were monitored over long periods of time and for these sources
the spectral parameters can be taken as good longterm average baselines for time-resolved spectral analysis.

\subsection{The Circumnuclear Material}

There are a number of factors that affect the shape and relative strength of the CRH: the photon index of the incident
power law, the inclination angle, the covering fraction of the material relative to the illuminating source, elemental abundances, 
and the geometry of the reflecting material.  Unfortunately, the changes in shape are very subtle and high energy spectrometers 
are not sensitive enough to these subtle differences to deconvolve all of these effects through spectral modeling.  
This has led to simplifications in the models and assumptions about the geometry of the reflecting material.
Common CRH models in use typically assume either a flat disk or a uniform torus, but since both produce such similar spectral signatures, 
we must use other techniques to discern the geometry of the Compton thick material.

The \pex model has been widely used to model reflection off a disk of Compton thick material such as the accretion disk.
It assumes a plane of Compton-thick material covering between 0 and $\sim$2$\pi$ sterradians of the sky from the point of view of the illuminating source 
corresponding to $R$ between 0 and $\sim$1.  However, a number of objects in our
sample have reflection fractions greater than 1, which is unphysical in this model, as is freezing the inclination angle
to 30$\degr$ for all objects.  It may seem tempting at this point to choose more accurate inclination angles on an object by object basis,
however accurate estimates are very difficult to come by.  It is sometimes assumed, based on Seyfert 1/2 unification schemes, that Seyfert 1's
will have smaller inclination angles (i.e.\ ``face on'' to the observer) while Seyfert 2's will have larger angles 
(i.e.\ ``edge on'' to the observer), but this poses a number of problems.  The first is that if we assume these objects have inherent differences
we will inevitably find that we are correct.  For example assuming Seyfert 2's are on average at an angle of $60\degr$ will inflate the value 
of $R$ by a factor of 1.2--1.5, leading to the possibly erroneous conclusion that there is more Compton thick material surrounding Seyfert 2's.

Additionally, this does not take into account what effect a torus may have if present.  
We attempted to apply the torus model \textsc{MYTorus} (Murphy \& Yaqoob 2009) to our data.
This model is a simple donut shape of uniform density with an opening angle of 60$\degr$.
Unfortunately, the model's assumption that the torus has a uniform density leads to
a steep change in the line-of-sight absorption at the edge of the donut-shaped torus, causing all Compton-thin
sources to have fitted inclination angles close to 60$\degr$.  Most sources also required additional Fe line emission from
Compton-thin material, which meant that the Fe line could not be used to constrain the amount of Compton-thick material in these sources.  
For the majority of our sources this led to two parameters, angle and torus density, to characterize only one measurable quantity: the flux of the CRH.

We concluded that the best way to proceed was with the results of the \pex model, but to utilize it in a predominantly phenomenological way.  
Since \pex is not the only CRH model available, we also report the flux ratios in Tables \ref{tabpar1} -- \ref{tabpc}, 
the relative flux of the CRH to the underlying power law near the peak of the CRH between 15 and 50 keV, which we can use to compare to 
results using other models, including ionized reflectors, such as \textsc{reflionx}, or torus models such as \textsc{MYTorus}.  
We calculated this flux ratio ($FR$) by finding the 15--50 keV flux for the power law continuum and for the CRH, then defining $FR=F_{\rm CRH} / F_{\rm Cont}$.
There is a linear proportionality between $FR$ and $R$ for fixed values of cos$i$ and $\Gamma$.  At 30$\degr$ with $\Gamma=1.9$, we find that $FR=0.8R$.
Patrick \etal (2012) reported $FR$ values for their 46 Seyfert 1's modeled with \textsc{reflionx}.
Comparing their distribution of reflection fractions (their Figure 12) with ours, shown in Figure \ref{FRdist}, we see a very similar smooth distribution 
with the majority of objects falling below $R$=1 ($FR$=0.8) but with a long tail towards higher values.

Using the Kolmogorov-Smirnov Test to compare our distributions of $R$ for Seyfert 1's versus Seyfert 2's we find a P value of 0.999 (P=0.982 comparing distributions of $FR$), 
where we have ignored the outliers NGC~6240, Circinus and NGC~4945, and included only well-determined $R$ values (i.e., with $\sigma_{R}/R > 1$ and upper limits $<$\,0.5). 
Here, $P$ denotes the likelihood that the two distributions of values of $R$ (or $FR$) can arise from the same parent population and thus confirms that the distributions are 
statistically similar between both classes of objects
These distributions are likely not consistent with the simple disk geometry and the standard Seyfert 1/2 unification since we would expect to see more Compton 
reflection in face-on Seyfert 1's than in side-view Seyfert 2's, which we do not observe.  The similarity in reflection fractions in our Seyfert 1's and 2's is consistent with
reflection off the inner wall of a torus, where viewing angle does not change the amount of observed reflection significantly.  We tested this empirically using the  \textsc{MYTorus}  and \pex models
and found a factor of $\lesssim$\,10\% difference in $FR$ going from 0\degr viewing angle to 60\degr for the torus, in contrast to a factor of $\sim$40\% difference over the same angle change for a disk.  
At viewing angles greater than $\sim$60\degr obscuration from the torus becomes a factor.

Another way to probe the circumnuclear material is to compare the strength of the Fe line with the strength of the Compton hump.  
George \& Fabian (1991) calculated the expected flux of the Fe line produced by reflection off a Compton-thick disk with 
respect to the flux of the Compton hump; it's expected that the Fe line EW will be $\sim$\,150 eV for $R \sim$\,1 at an angle of 30 $\degr$.  
In Figure \ref{RvsEW} we show the strength of the CRH plotted versus the EW of the Fe line for all Seyferts. 
From our analysis, Compton-thick reflection accounts for $\sim$\,40\% of the Fe line flux on average, however the variation is quite large.  
The remainder of the Fe line flux may arise in Compton-thin neutral gas in the NLR or in the BLR clouds
(although there is a limit to how much Fe line flux Compton-thin gas can produce; see De Rosa \etal 2012) and/or 
be from ionized Compton-thin gas in the vicinity of the nucleus. 
With the limited resolution of the PCA around 6--7 keV we cannot disentangle these possible origins.
Approximately three quarters of Seyfert 1's and Compton-thin 2's show more Fe line flux from sources other than from 
Compton reflection, assuming a disk geometry.  Note that Compton-reflection from a torus may have a slightly different 
expected ratio of the Fe line to the CRH flux, depending on the particular geometry assumed.

\begin{figure}
  \plotone{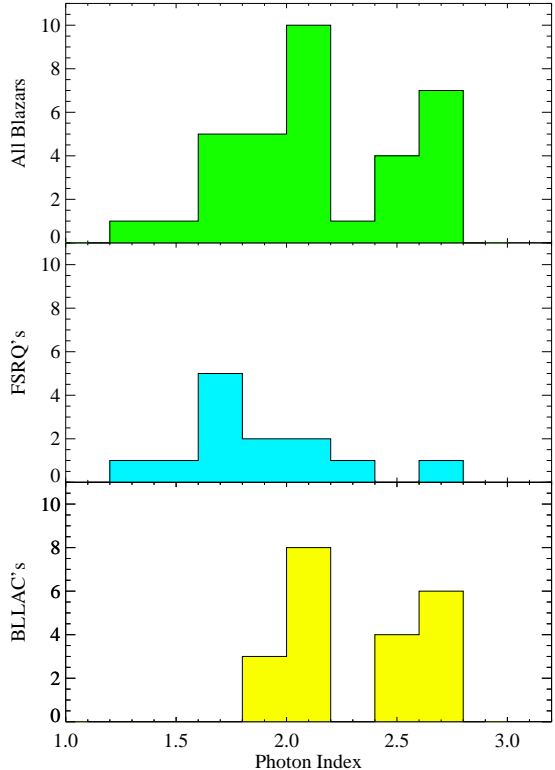}
  \caption{$\Gamma$ distribution by blazar type.  BL~Lac objects tend to have higher photon indices than FSRQ's.}
  \label{Gdistbl}
\end{figure}

\begin{figure}
  \plotone{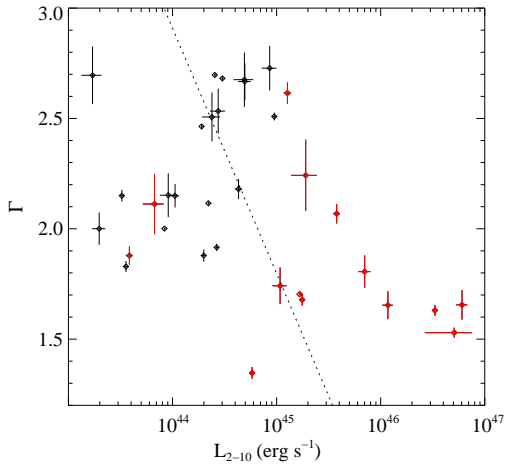}
  \caption{Photon index versus 2--10 keV luminosity for the blazars in our sample.  Red data points are FSRQ's; black data points are BL~Lac objects. 
           We find significant negative correlation which is consistent with the Fossati sequence. }
  \label{LGblaz}
\end{figure}
\subsection{Results for the Blazar Sample}

Most of the blazars in our sample were well fit by a simple power law.  
Four BL~Lac type objects were fit better by a broken power law with break energies below $\sim$\,10 keV and steepening by $\sim$\,0.2
(note that this small a change in $\Gamma$ is due to a very gradual rollover of the spectrum which we are only sensitive to in our brightest sources).  
The average $\Gamma$ for our sample was 2.1 with a variance of 0.15.  The average $\Gamma$ for
flat spectrum radio quasars (FSRQ's) was 1.8 while the average for BL~Lac objects was 2.3 with variances of 0.2 and 0.1 respectively.
The distribution of $\Gamma$ for the blazar sub-types is shown in Figure \ref{Gdistbl}.

In addition to the lower photon index, the FSRQ's also tend to be much more luminous than the BL~Lac's.
Figure \ref{LGblaz} shows the X-ray 2--10 keV luminosity versus the photon index for our sample of blazars.
We found a weak anti-correlation between these quantities with a Pearson correlation coefficient of --0.40, significant
at the 99\% level.  This is consistent with the ``Fossati sequence'' (Fossati \etal 1998) which predicts that
for higher luminosities the peak of the broadband emission humps would shift to lower energies.  
At lower luminosities the upper end of the synchrotron hump dominates the X-ray band 
while at higher luminosities the lower end of the inverse Compton hump dominates.
The means a softening of the X-ray portion of the SED and an increase in the photon index as luminosity decreases.  
Donato \etal (2001) published a large X-ray sample of blazars observed by \sax which also confirmed this trend.

Dai \& Zhang (2003) analyzed {\it EXOSAT}, {\it ASCA} and \sax data to obtain hard X-ray photon indices for 20 sources.
They reported average $\Gamma$ of 2.78$\,\pm\,$0.25 for high-frequency-peaked BL~Lac objects (HBLs),
1.85$\,\pm\,$0.35 for low-frequency-peaked BL~Lac objects (LBLs), and 1.64$\,\pm\,$0.28 for FSRQs.
Fan \etal 2012 focused on Fermi-selected blazars, finding average values of $\Gamma$ in the X-ray band of 2.39$\,\pm\,$0.35 for HBLs,
1.97$\,\pm\,$0.38 for LBLs, and 1.89$\,\pm\,$0.37 for FSRQs.  Noting that the majority of our BL~Lac objects are 
high-frequency or intermediate-frequency peaked, these measurements are all in good agreement with ours.

\subsection {Conclusion}

We have analyzed data for 100 AGN in the \xte archive in order to explore the geometry of circumnuclear material around SMBH's and 
characterize their X-ray spectra. 
We present a large sample of X-ray bright AGN which has a number of advantages over previous surveys in the medium--hard X-ray bandpass:
the simultaneity of the $<$10 keV and $>$10 keV X-ray data eliminates ambiguity, the relatively 
broad bandpass is necessary to accurately measure the continuum power law and the CRH in order to gain insight into the 
geometry and characteristics of the circumnuclear material, and the long monitoring campaigns make this sample ideal
for using as a baseline average for future time-resolved spectral analysis.
To that end, we have presented the spectral parameters of our sample including absorption, Fe line equivalent widths, Compton reflection 
strengths, photon indices, and 2--10 keV fluxes in Tables \ref{tabpar1}--\ref{tabbp} along with average
values by type in Table \ref{tabavg}.

The similar distributions of $\Gamma$ for type 1 and 2 Seyferts supports the idea that they share a common central engine.
The distribution of $R$ showed no difference in the reflection from Compton-thick material in Seyfert 1's and 2's.
This is counter to what we would expect from reflection off a disk under classical unification schemes where face-on Seyfert 1's would be expected to
show significantly more reflection than edge-on Seyfert 2's.  The similar distributions are more consistent with reflection off a torus.
We did not find a significant correlation between $\Gamma$ and $L_{2-10}$ for the Seyferts in our sample, however NLSy1's showed significantly
higher photon indices.  This is consistent with a common central engine for all Seyferts with the primary differences between types
being dependent on accretion rate and the geometry of the circumnuclear material.  
Additionally, a few of the Seyfert 2's in our sample had very soft X-ray spectra, similar to NLSy1's, and could be type 2 analogs of NLSy1's, as expected under unification.

We found that roughly 85\% of Seyferts showed significant contribution from the CRH.
Comparing the strength of the CRH with the amount of the Fe emission seen allowed us to estimate the ratio of Compton-thick 
to Compton-thin material in AGN with the average being around 40\%, however with large object to object variation.

We found a negative correlation between $\Gamma$ and $L_{2-10}$ for blazars in agreement with the Fossati sequence and the 
luminosity dependance of the broad band SED hump peak energies.

We confirm that it is likely AGN do share a common engine across the various types, concluding that the differences in their observed properties
are likely based on mass, accretion rate, and geometry of the circumnuclear material.  The ratio of Compton-thick to Compton-thin material was not
consistent from object to object and did not seem to be dependent on optical classification.  While Seyfert 2's were more likely to have high absorption columns,
they were as likely to show strong Compton reflection humps as Seyfert 1's, inconsistent with reflection off a disk, assuming type depends on inclination angle.   
A more complex reflecting geometry such as a torus, combined disk and torus, or clumpy torus is likely a more accurate picture of the Compton-thick material.

As a resource for future analyses of AGN, we plan to make spectra and light curves available online.  
These spectra may serve as baselines for future missions such as \textsl{NuSTAR}, which which will look at a large number of AGN in the 6--80 keV band with excellent sensitivity and spectral resolution.
They could potentially be combined with multi-wavelength data to create SED's for future analysis or with information about abundances of the different types to create the AGN contribution to the CXB.

\begin{acknowledgments}
The authors would like to thank Joern Wilms and Craig Markwardt for valuable advice, 
as well as the anonymous referee for their diligence and patience.
This research has made use of data obtained from the \textsl{RXTE} satellite, a NASA space mission.
This work has made use of HEASARC online services, supported by NASA/GSFC, and the NASA/IPAC Extragalactic Database, 
operated by JPL/California Institute of Technology under contract with NASA.
The research was supported by NASA Contract NAS 5-30720 and Grants NNX09AG79G and NNX11AD07G.
\end{acknowledgments}


\appendix
\section{A. Rejected Sources}

Approximately 60 AGN were observed in the lifetime of \xte which are not included in our sample.  
Most of them were very faint and/or were not observed for very long.  
A few objects were contaminated by other X-ray bright sources in the field of view. For example, 3C 84 is known to be embedded in an X-ray
bright galaxy cluster (e.g., Fabian et al.\ 2006) whose emission dominated the PCA spectrum.  Observations of NGC 6814 have the
cataclysmic variable V1432 Aql in the field of view (Mukai et al.\ 2003).  Observations of the blazar RGB J1217+301 were consistent with
detecting only contaminating flux from the NLSy1 Mkn 766 and the blazar 1ES 1218+304, located approximately 0.33 and 0.76 degrees away, respectively.
The remaining AGN are listed in Table \ref{tabrej} including their type as determined by NED, PCA exposure time, $F_{2-10}$, and $\Gamma$ where it was possible to constrain.

\begin{deluxetable*}{llccc}
   \tablecaption{Other AGN in the \xte Archive\label{tabrej}}
   \tablecolumns{5}
\startdata
\hline
\hline\\[-2mm]
Source      &               &    PCA Exposure     &     Flux$_{2-10}$                         & \\
Name        &      Type     &    (ks)          &     ($10^{-12}$ \fluxunits)  &   $\Gamma$ \\[1mm]
\hline\\[-2mm]
Seyfert 1's\\[1mm]
\hline\\[-2mm]
H\,0147--537                &   QSO         & 83.3 		&   2.7$\,\pm\,$0.1  &  1.9$\,\pm\,$0.2 \\
1H\,0707--495          	&   NLSy1     & 7.8 		&   4.3$\,\pm\,$1.7  &  3.5$\,\pm\,$0.5 \\
LBQS 2212--1759    &  BALQSO  &    18.3      	&  $1.4^{+0.1}_{-0.4}$  &     $2.3\pm0.5$ \\
PG\,1116+215         	&  Sy1     	    &    51.1    	&  3.5$\,\pm\,$0.1   &   1.83$\,\pm\,$0.11 \\
PG\,1416--129       	&   Sy1  	    & 22.4 		&  3.7$\,\pm\,$2.4  &  1.5$\,\pm\,$0.3 \\
PG\,1440+356 (Mkn\,478) & NLSy1 &   27.9      &  $3.2^{+0.1}_{-0.3}$  &     $2.8 \pm 0.3$ \\
PG\,1700+518        	&  Sy1/BALQSO  &  27.4   &   $< 0.2$   &    -         \\
RHS\,03              		&   Sy1    	 & 7.0 		&   8.2$\,\pm\,$  6.3  &  1.9$\,\pm\,$0.2 \\
RHS\,15              		&  Sy1     	 &   9.8     		&   $2.6\pm0.2$            &   $1.0 \pm 0.2$ \\
RHS\,17              		&   Sy1    	 & 9.9 		&   7.5$\,\pm\,$  1.6  &  1.6$\,\pm\,$0.2 \\
RHS\,54              		&  Sy1     	 &   7.4     		&   1.8\err{0.2}{0.6}     &  1.2$\,\pm\,$0.5 \\
RHS\,56              		&   NLSy1  & 10.2 		&   5.8$\,\pm\,$0.8  &  2.2$\,\pm\,$0.2 \\
RHS\,61              		&   Sy1    	 & 8.9 		&   4.8$\,\pm\,$1.7  &  1.9$\,\pm\,$0.3 \\
TON1542 (Mkn\,771)  & Sy1      &    90.9     	&   $3.8\pm0.1$         &      $2.0\pm0.1$ \\[1mm]
\hline\\[-2mm]
Seyfert 2's \\[1mm]
\hline\\[-2mm]
Arp 220            & Sy2/ULIRG &   0.9     &   $<1.5$           &               -  \\
E 253-G3           &     Sy2   & 1.7 &   3.7$\,\pm\,$ 3.7  &  0.9$\,\pm\,$0.6 \\
IRAS\,F00521--7054 & Sy2       &    1.5     &   $4.2^{+0.4}_{-2.1}$   &           1.6$\,\pm\,$0.5 \\
IRAS\,F01475--0740 & Sy2/ULIRG  &   3.0     &   $<1.3$           &               -  \\
IRAS\,F03362--1642 & Sy2        &   1.8     &   $<1.3$           &               -  \\
IRAS\,F04385--0828 & Sy2        &   1.9     &   $4.7^{+0.5}_{-2.8}$   &           1.6$\,\pm\,$0.4 \\
IRAS\,F05189--2524 & Sy2/ULIRG  &   1.4     &   $<3.3$         &                 $1.9^{+0.9}_{-0.8}$ \\
IRAS\,F08572+3915  & Sy2/ULIRG   &  3.4     &   $<2.7$         &                  1.9$\,\pm\,$0.7 \\
IRAS\,F19254--7245 (AM 1925--724) & Sy2 &  1.8     &   $<1.5$         &       -      \\              
MCG--3-34-63       &    Sy2     & 1.8 &   1.8$\,\pm\,$1.8  &  -   \\
NGC\,1320           & Sy2         &  3.4     &   $<0.9$        &             -  \\
NGC\,1386            &    Sy2     & 2.5 &   5.8$\,\pm\,$2.3  &  2.7$\,\pm\,$1.0 \\
NGC\,3281           & Sy2/C-thick & 11.4 &   8.0$\,\pm\,$2.4  &  2.6$\,\pm\,$1.5 \\
NGC\,3660           & Sy2      &   3.2     &   $<2.0$        &   1.8\err{0.9}{0.8}          \\ 
NGC\,5347           & Sy2        &   3.2     &   $<1.8$        &             $1.7^{+1.0}_{-0.9}$ \\
NGC\,6251          &   Sy2       &  148    &   3.2\err{0.1}{0.4}  & 2.38\,$\pm\,$0.23  \\[1mm]
NGC\,6394          &  Sy2        &  23.8     &   $<1.9$        &            -  \\
NGC\,6890          &  Sy2      &   2.5     &   $1.1\pm0.3$   &             0.9$\,\pm\,$1.3 \\
TOL\,1238--364 (IC 3639)  &  Sy2   &   1.7   &  $<1.4$  &                   $2.9^{+3.1}_{-1.7}$ \\[1mm]
\hline\\[-2mm]
Blazars\\[1mm]
\hline\\[-2mm]
0420--014   &      BLLAC       &   1.0    &    <2.8       &   -    \\
1ES\,0806+524     &    BLLAC  &     39.4        &   5.6\err{0.3}{0.8}    &   2.8$\,\pm\,$0.2      \\
3C\,446     &       BLLAC      &  40.5       &      <1.5       &      2.0$\,\pm\,$0.6         \\
4C\,38.41        &   FSRQ          &    65.5  &    1.7\err{0.2}{0.5} &  1.45$\,\pm\,$0.30 \\
H\,2356--309     &    BLLAC         & 2.2          &    8\err{2}{7}        &    2.4$\,\pm\,$0.5     \\
O\,J287            &     BLLAC    &    116.8    &       $<0.2$  &  -  \\
PG\,1424+240    &       BLLAC         & 32.0          &    2.6\err{0.1}{0.4}        &    3.6$\,\pm\,$0.5     \\
PKS\, 0235+164     &     BLLAC    &   247.5       &     1.6\err{0.3}{0.8}     &    2.5$\,\pm\,$0.3     \\
PKS\, 0332--403  &     FSRQ         &  18.8     &    3.6\err{0.3}{0.6}       &   2.5$\,\pm\,$0.4     \\
PKS\, 0348--120     &    FSRQ    &      93.6    &       $< 0.1$  &  -  \\
PKS\, 0405--385    &     FSRQ    &   1.9    &       $< 0.2$  &  -  \\
PKS\,0537--286   &        QSO     &   23.7    &     4.1\err{0.1}{1.3}      &   1.3$\,\pm\,$0.2    \\
PKS\,0537--441   &    BLLAC         &   19.3    &     3.6\err{0.4}{1.5}      &   2.8$\,\pm\,$0.5    \\
PKS\,2255--282   &     FSRQ        &   5.4     &     8.0\err{0.3}{0.7}      &   1.67$\,\pm\,$0.14    \\
RGB\,J0152+017   &     BLLAC        &   31.1    &   6.4$\,\pm\,$0.4        &   2.47\err{0.19}{0.18}    \\
RHS\,53                &    BLLAC         &  9.2     &      3.7\err{0.3}{0.5}     &    1.74$\,\pm\,$0.23    \\
W Com (RGB\,J1221+282)  &    BLLAC     &     23.7     &     $<1.8$ &    $1.7\pm0.7$ \\
\enddata
\tablecomments{\xte archival AGN which were not included in our main sample with object NED type, PCA exposure time, 2--10 keV flux and $\Gamma$ where it could be constrained.  Where $\Gamma$ could not be constrained a photon index of 2.0 was assumed to find the upper limit to the flux.  Note that the errors given are purely statistical and do not reflect systematic uncertainties in the background.  Thus some of these sources may seem to have $\Gamma$ constrained to within 10\% such that they could be included in the sample, however due to low flux or very short exposure the background systematics are large enough that they were not included. ``-'' indicates an unconstrained parameter.  ULIRG is an ultra luminous infrared galaxy and BALQSO is a broad absorption line quasar.}
\end{deluxetable*}

\section{B. Notes on Individual Sources}

Several sources in our sample required complex modeling or extra analysis.  
Many Compton-thick sources showed evidence for a soft power law component lacking intrinsic absorption.  
We have searched through the literature to find explanations for the spectral characteristics of each of these sources.  

Circinus is a bright, well-studied, reflection-dominated source with a very strong CRH, a Compton-thick absorber, and a 
soft power law component, the origin of which is a combination of ionized plasma commensurate with the NLR and contamination from nearby point sources (Matt \etal 1999; Sambruna \etal 2001).  
We have measured a high energy rollover in this source around $\sim$40 keV, consistent with that of $\sim$50 keV found by Yang \etal (2009) using \suzaku.

The soft power law in Mkn\,3 has been identified by Chandra (Sako \etal 2000) 
and confirmed with \xmm (Bianchi \etal 2005) as originating in photoionized plasma in the NLR.

NGC\,1068 is a very weak, reflection dominated source and was not observed for only 54 ks by \xte, making it difficult to properly constrain the complex scenarios 
that have been modeled previously in this source. It was possible to fit this source with with a power law continuum plus a CRH (Table \ref{tabpar2}) 
or with a Compton-thick absorber with leaked emission and a CRH (Table \ref{tabpc}).  
Matt \etal (1997) and Colbert \etal (2002) modeled the spectrum below 10 keV with an ionized reflector plus a neutral reflector, 
however because the source is so faint we cannot distinguish between this and a power law.
Since we are unable to place constraints on the Compton-thick absorber in this source due to the extremely high column density and lack of good data above $\sim\,$30 keV,
we adopt the parameters from the base model given in Table \ref{tabpar2} for this source.

NGC\,4945 was best fit by a hard X-ray power law with a Compton-thick absorber and an additional power law visible 
below about 10 keV due to nuclear starburst activity (see, e.g., Schurch \etal 2002 and references therein).
NGC\,4945 required a high energy rollover at $\sim\,$60 keV for a good fit (see RMR2011 for more details on NGC\,4945).

NGC\,6240 has scattered nuclear emission below $\sim\,$10 keV found by Lira \etal (2002) using \chandra,
with possible starburst contamination and a binary nucleus (Komossa \etal 2003).

NGC\,6300 has been reported to have very interesting spectral behavior, changing from a reflection dominated to a regular Compton-thin Seyfert 2 
(Leighly \etal 1999; Guainazzi 2002), while a variability study by Awaki \etal (2005) has indicated that it may be a Seyfert 1 core obscured by heavy absorption.
However, the faintness of the source means that it has been difficult to properly constrain these models in order to test these ideas.
With only 27 ks of data and a flux of $\sim6 \times 10^{-12}$ \fluxunits, we are not able to constrain these models very well either.
Fitting the data with a reflection-only model yields \chidof $\sim 1$, and $\Gamma \sim 2.1$, 
while a partial-covering Compton-thick absorber model with no reflection yields \NH $\sim3 \times 10^{24}$ cm\e{-2} and \chidof $\sim 0.5$.
Neither of these is an improvement over our base model fit, but are plausible fits to the data.

NGC\,7582 may also be reflection dominated, however it was not possible to disentangle the absorbed power law and the CRH for these sources.  
NGC\,7582 has been modeled with very strong neutral reflection plus an underlying power or with a partial covering absorber of the continuum 
(Turner \etal 2000; Bianchi \etal 2007) and we find that modeling either of these components provided a good fit to our data. 
Note that if \xte caught the source sometimes Compton-thick and sometimes Compton-thin (Piconcelli \etal 2007, Bianchi \etal 2009), 
then a partial covering absorber model would provide a good fit of the time-averaged data and would make it difficult to measure accurately the actual quantity of reflected emission.

For almost all of the blazars in our sample a simple or broken power law gave a good description of the spectrum, however 3C\,273 is something of a 
special case with an Fe line and Compton hump having been detected previously in this source (e.g., Grandi et al.\ 1997, Kataoka et al.\ 2002).
A Compton reflection component was significant to include at the 96\% confidence level but with a very low value of $R$ (0.04$\pm\,$0.03) and
it was significant to include an Fe line at only the 86\% confidence level with an EW of 50$\,\pm\,$40 eV.
This is consistent with a Seyfert-like source being diluted by the jet component in the X-ray band.

The blazars 1ES 1218+304 and Mkn~766 are located 0.73 degrees from each other.  
We therefore performed additional analysis to determine the level of contamination in each source.
At this off-axis angle, the PCA response is $\sim21\%$.  We measure a 2--10 keV flux for Mkn 766 of $2.7\times 10^{-11}$ erg cm$^{-2}$s$^{-1}$, 
consistent with measurements many other X-ray missions (e.g., Ueda et al.\ 2001, Markowitz et al.\ 2007).
Meanwhile, we measure a 2--10 keV flux for 1ES 1218+304 of $1.2\times 10^{-11}$ erg cm$^{-2}$ s$^{-1}$; previous 
2--10 keV flux measurements span the range 1.5 to $2.6\times 10^{-11}$ erg cm$^{-2}$ s$^{-1}$ 
(from \textit{BeppoSAX}: Giommi et al.\ 2005; \textit{XMM-Newton}: Blustin et al.\ 2004; \textit{Suzaku}: Sato et al.\ 2008).
It is likely that the 1ES 1218+304 spectrum is contaminated at some level, but given the high variability of Mkn~766 we can 
only give a rough estimate of the exact amount of contaminating flux.  \sax observed Mkn~766 in May of 1997, $\sim$10 days before 
the \xte observations of 1ES 1218+304, and measured the 2--10 keV flux to be $2.0 \times 10^{-11}$\fluxunits (Matt \etal 2000).
This would correspond to a contaminating flux of $\sim 0.4 \times 10^{-11}$\fluxunits in the 2--10 keV range, introducing systematic 
errors on the measurement of the photon index of $\sim$0.1 and on the overall flux at around the 30\% level.

\end{document}